\documentclass[12pt,preprint]{aastex}

\newcommand{\asca}{{\it ASCA} }

\newcommand{\ginga}{{\it Ginga} }

\newcommand{\rxte}{{\it RXTE} }
\newcommand{\bepposax}{{\it BeppoSAX} }

\newcommand{\bq}[1]{\begin{equation} \label{#1}}
\newcommand{\eq}{\end{equation}}

\shortauthors{Kataoka et al.}
\shorttitle{Characteristic X-ray Variability of TeV Blazars}

\begin{document}

\title{Characteristic X-ray Variability of TeV Blazars:  \\
   Probing the Link between the Jet and the Central Engine}

\author{Jun Kataoka\altaffilmark{1}, Tadayuki Takahashi\altaffilmark{2}, 
Stefan J. Wagner\altaffilmark{3}\\
Naoko Iyomoto\altaffilmark{2}, Philip G. Edwards\altaffilmark{2},  
Kiyoshi Hayashida\altaffilmark{4},\\
Susumu Inoue\altaffilmark{5}, Greg M. Madejski\altaffilmark{6}, 
Fumio Takahara\altaffilmark{4},\\
 Chiharu Tanihata\altaffilmark{2}, and Nobuyuki Kawai\altaffilmark{1}} 
\altaffiltext{1}{Department of Physics, Faculty of Science, Tokyo
Institute of Technology, Megro-ku, Tokyo, 152-8551, Japan}
\altaffiltext{2}{Institute of Space and Astronautical Science,
Sagamihara, Kanagawa 229-8510, Japan}
\altaffiltext{3}{Landessternwarte Heidelberg, K\H{o}nigstuhl, D-69117
Heidelberg, Germany}
\altaffiltext{4}{Department of Earth and Space Science, Graduate School of
Science, Osaka University, Toyonaka, Osaka, 560-0043, Japan}
\altaffiltext{5}{Theoretical Astrophysics Division, National Astronomical
Observatory, Mitaka, Tokyo 181-8588, Japan}
\altaffiltext{6}{Stanford Linear Accelerator Center, 2575 Sand Hill Rd. 
Menlo Park, CA, 94025}

\begin{abstract}

We have studied the rapid X-ray variability of three extragalactic 
TeV gamma-ray sources;
Mrk~421, Mrk~501 and PKS~2155$-$304. Analyzing the X-ray light 
curves obtained from \asca and/or \rxte observations 
between 1993 and 1998, we have 
investigated the variability in the time domain from 10$^3$ to 10$^8$ sec. 
For all three sources, both the power spectrum density 
(PSD) and the structure function (SF) show a roll-over with a time-scale
of the order of 1~day or longer, which may be interpreted as the typical 
time-scale of successive flare 
events. Although the exact shape of turn-over is not well constrained and the 
low-frequency (long time-scale) behavior is still unclear, the 
high-frequency (short time-scale) behavior is clearly resolved.  
We found that, on time-scales shorter than 1~day, there is 
only small power in the variability, as indicated by a steep power 
spectrum density of $f^{-2 \sim -3}$. This is very different 
from other types of mass-accreting black-hole systems for which the short 
time-scale variability is well characterized by a fractal, flickering-noise 
PSD ($f^{-1 \sim -2}$). 
The steep PSD index and the characteristic 
time-scale of flares imply that the X-ray emitting site in the 
jet is of limited spatial extent; $D$ $\ge$ 10$^{17}$~cm distant from 
the base of the jet, which corresponds to $\ge$ 10$^{2}$ Schwarzschild 
radii for 10$^{7-10}$ $M_\odot$ black-hole systems. 

\end{abstract}


\section{Introduction}

Observations with the EGRET instrument (30 MeV$-$30 GeV; Thompson et 
al.\ 1993) on-board the {\sl Compton Gamma-Ray Observatory} ({\sl CGRO}) 
have detected $\gamma$-ray emission from over 60 AGNs
(e.g., Hartman et al.\ 1999). 
Most of the AGNs detected by EGRET show characteristics of the
blazar class of AGN -- such as violent optical flaring, high optical
polarization, and flat radio spectra (Angel and Stockman 1980).
Observations with ground-based Cherenkov telescopes have 
detected $\gamma$-ray emission extending up to TeV energies for a
number of nearby AGN. We limit our discussion in this paper to the three 
TeV sources for which substantial X-ray data-sets are available:
Mrk~421 ($z$ = 0.031; TeV detection reported by Punch et al.\ 1992), 
Mrk~501 ($z$ = 0.034; Quinn et al.\ 1996), 
and PKS 2155$-$304 ($z$ = 0.117; Chadwick et al.\ 1999). 

The overall spectra of blazars (plotted as $\nu$$F_{\nu}$) have two 
pronounced continuum components: one peaking between IR and X-rays, and 
the other in the $\gamma$-ray regime (e.g., von Montigny et al.\ 1995). 
The lower energy component is believed to be produced by 
synchrotron radiation from relativistic electrons in magnetic fields, 
while inverse-Compton 
scattering by the same electrons is thought to be the dominant process 
responsible for the high energy $\gamma$--ray emission 
(e.g., Ulrich, Maraschi, \& Urry 1997). The radiation is emitted from 
a relativistic jet, pointing close to our line of sight 
(e.g., Urry \& Padovani 1995).  VLBI observations of superluminal 
motions confirm that the jet plasma is moving 
with Lorentz factors of $\Gamma$ $\simeq$ 10 (e.g., Vermeulen \& Cohen 1994).

Blazars are commonly variable from radio to $\gamma$-rays.
The variability time-scale is shortened and the radiation is strongly
enhanced by relativistic beaming. For extragalactic
TeV sources, the X-ray/TeV $\gamma$-ray bands correspond to the highest 
energy ends of the synchrotron/inverse-Compton emission,
which are produced by electrons accelerated up to the maximum energy 
(e.g., Inoue \& Takahara 1996; Kirk, Rieger \& Mastichiadis 1998; 
Kusunose, Takahara \& Li 2000). At the highest energy ends, variability 
is expected to be most pronounced; indeed, multi-frequency campaigns of 
Mrk~421 have reported more rapid, and larger amplitude variability 
in both X-ray and TeV $\gamma$-ray bands than those in other wavelengths 
(e.g., Macomb et al.\ 1995; Buckley et al.\ 1996; 
Takahashi et al.\ 1999; 2000). Thus the X-ray variability can be the most 
direct way to probe the dynamics operating in jet plasma, in particular 
compact regions of shock acceleration which are presumably 
close to the central engine.

`Snapshot' multiwavelength spectra principally provide us with clues
on the emission mechanisms and physical parameters inside relativistic jets.
On the other hand, detailed studies of time variability
not only lead to complementary information for the objectives above,
but should also offer us 
a more direct window on the physical processes operating in the jet
as well as on the dynamics the jet itself. 
However, short time-coverage and under-sampling have 
prevented detailed temporal studies of blazars. 
Only a few such studies have been made in the past for blazars, 
e.g., evaluation of the energy dependent 
``time-lags'' based on the synchrotron cooling picture 
(e.g., Takahashi et al.\ 1996; Kataoka et al.\ 2000; Takahashi et al.\ 2000).

Variability studies covering a large dynamic range 
and broad span of time-scales have become common
for Seyfert galaxies and Galactic black-holes (e.g., Hayashida et al.\ 1998; 
Edelson \& Nandra\ 1999; Chiang et al.\ 2000). From 
power spectrum density (PSD) analyses, it is well known that 
rapid fluctuations with frequency dependences
$P(f)$ $\propto$ $f^{-1 \sim -2}$, are characteristic of time variability 
in accreting black hole systems. Although their physical origin is still 
under debate, some tentative scenarios have been suggested 
to account for these generic, fractal 
features (e.g., Kawaguchi et al.\ 2000).  

Our main goal is to delineate the characteristic time variability of the 
X-ray emission from TeV gamma-ray sources, highlighting 
the differences between 
such $jet$-$enhanced$ objects (blazars) and sources without prominent jets 
(Seyfert galaxies and Galactic black-holes).  Very recently, three TeV sources 
were intensively monitored in the X-ray band (Kataoka 2000; Takahashi 
et al.\ 2000), providing valuable information on the temporal behavior of 
these objects. The most remarkable result is a detection of a clear 
``roll-over'' in the structure function (SF; $\S$~2.4) of Mrk~421 in a 1998 
observation  (Takahashi et al.\ 2000). Such a roll-over, if confirmed,  would 
yield considerable scientific fruits about the physical origin and the 
location of X-ray emission inside the jet plasma, which is the primary 
motivation of 
this paper. Systematic studies using a larger sample of data will
be necessary to confirm and unify the variability features in blazars. 

Combining data from archival \asca and \rxte observations over 5 years 
(3 years for \rxte), we derive here the variability information on time-scales 
from minutes to years.  This is the first report of variability analysis 
of blazars based on high quality data covering the longest observation 
periods available at X-ray energies.
The observations and data reduction are 
described in $\S$~2.1 and $\S$~2.2. Temporal studies using the PSD are
described in $\S$~2.3, while alternative approach using the SF is considered
in $\S$~2.4. In $\S$~3, we discuss the origin of the rapid variability. 
Finally, in $\S$~4 we present our conclusions.
 
\section{Temporal Analysis}

\subsection{Observations}

The three extragalactic TeV sources were observed a number of times with the 
X-ray satellites \asca and/or \rxte. Observation logs are given in
Table~1 and 2. \asca observed Mrk~421 five times with a net exposure 
of 546~ksec between 1993 and 1998.  In the 1998 observation, the source was in
a very active state and was detected at its highest-ever level
(Takahashi et al.\ 2000). \rxte observed Mrk~501 more than 100 times with 
a net exposure of 700~ksec between 1996 and 1998.  Mrk~501 was in a
historical high-state in 1997 (Catanese et al.\ 1997; Pian et al.\ 1998; 
Lamer \& Wagner 1999). Multiwavelength campaigns, including a number of ToO 
(Target of Opportunity) observations, were conducted during this high state.  
PKS 2155$-$304 was 
observed with \asca for  133~ksec in 1993$-$1994, while 
observations  over 246~ksec were conducted with \rxte in 1996$-$1998. 
Full descriptions of the \asca and \rxte monitorings of TeV sources, 
other than those observations discussed in this paper, 
are given in Kataoka (2000).  In the following, we  classify the X-ray 
observations into two convenient groups based on their sampling strategy.  

The first group is the $continuous$ observations of more than 1 day, which 
enables detailed monitoring of the time evolution of blazars (Table~1).  
In particular, three `long-look' observations have been conducted; for Mrk~421 
(7 days in 1998), Mrk~501 (14 days in 1998), and PKS~2155$-$304 
(12 days in 1996).  For this group, the observing efficiency, which is 
defined as the ratio of net exposure to the observing time, is about 0.5 
for the \asca observation. Interruptions are due to Earth occultation and 
passages through South Atlantic Anomaly, etc. For the two \rxte observations, 
the observing efficiency was less; 0.3 for the 14 day monitoring of 
Mrk~501 and 0.2 for the 12 day monitoring of PKS~2155$-$304, 
respectively. We use the data from these $continuous$ observations in both 
the PSD ($\S$~2.3) and the SF analysis ($\S$~2.4).

The second group is the $short$ observations of a few ksec each, which are 
spaced typically $\ge$~1~day apart, so that the source can be monitored
over as long a time range as possible (Table~2). For these observations, 
the observing 
efficiency is less than 0.1 and/or there are large gaps during the 
observation. 
These interrupted observations are not suitable for 
the PSD studies described in $\S$~2.3, but are still useful for 
investigating the 
long-term variability based on the SF analysis ($\S$~2.4).

\subsection{Data Reduction}

All the \asca observations listed in Table~1 and Table~2 were performed 
in a normal PH mode for the Gas Imaging Spectrometer 
(GIS: Ohashi et al.\ 1996). A normal 4~CCD mode was used for the Solid-state 
Imaging  Spectrometer (SIS: Burke et al.\ 1991; Yamashita et al.\ 1997) 
for 1993 observations, while a normal 1~CCD mode was used for the 
observations after 1994. Standard screening procedures were applied to the 
data and the source counts are extracted from a circular region centered on 
a target with a radius of 6~arcmin for the GIS and 3~arcmin for the SIS. 
The count rate of both GIS detectors (GIS~2/3) and SIS detectors (SIS~0/1) 
are separately summed. Since the count rate of the background 
($\sim$~0.01~cts/s) and its fluctuation are negligible compared with the 
source count rates ($\ge$~1~cts/s), background subtraction was not performed. 

For the PKS~2155$-$304 observation in 1993, we only analyzed the 
GIS data because the source was so bright that the SIS detector was strongly 
saturated in 4~CCD mode. Similarly, for the Mrk~421 observation in 
1998 when the source was in an historically high state (Takahashi et al.\ 2000),
both the GIS and SIS were saturated during the observation.  
For this observation, we estimate the GIS count rate from the relation between 
the GIS count rate and the ``hit count'' of the Lower Discriminator 
(Makishima et al.\ 1996). The effects caused by saturation of the SIS 
detectors were corrected by extracting the source counts from a narrower 
circular region than usual; a radius of 1~arcmin for SIS0 and 2.6~arcmin for 
SIS1, respectively. 

For the \rxte observations, source counts from the Proportional Counter 
Array (PCA: Jahoda et al.\ 1996) were extracted from 3 Proportional Counter 
Units (PCU 0/1/2) which had much larger and less interrupted exposures than 
those for PCU3 and PCU4.  We used only signals from the top layer 
(X1L and X1R) in order to obtain the best signal-to-noise ratio. 
Standard screening procedures were performed on the data. 
Backgrounds were estimated using $pcabackest$ (Version 2.1b) for the 
PCA data. Although there are two other instruments on-board \rxte, the
High Energy X-ray Timing Experiment (HEXTE) and All-Sky Monitor (ASM), 
we do not use data from these instruments for several reasons. 
Firstly, calibration problems make the analysis 
results quite uncertain for both HEXTE and ASM. Secondly, the typical 
exposure for \rxte observation  was too short to yield statistically 
meaningful hard X-ray data above 20~keV for HEXTE.

In order to obtain the maximum photon statistics and the best signal-to-noise 
ratio, we selected the energy range 0.7$-$10 keV for the \asca GIS,
0.5$-$10 keV for the \asca SIS, and 2.5$-$20 keV for the \rxte PCA. 
The light curves for the $continuous$ \asca(GIS)/\rxte(PCA) observations 
(Table~1) are shown in Figure~1 (a)$-$(g). 
We plot the GIS data rather than the SIS data to compare the source count 
rate, because the GIS has a wider field of view than the SIS and is less 
affected by the attitude of the satellite and position of the source 
on the detector.  The binning time is 256~sec for 
\asca (GIS) data and 5760~sec for the \rxte (PCA) observations.
Note that 5760~sec is the orbital period of both \asca and \rxte satellites. 
Expanded plots of the \rxte (PCA) light curves are given in small panels in 
Figure~1 (d) and (g), with a binning time of 256 sec.  
Variability is detected in all of the observations. 
In particular, for the long-look monitoring of Mrk~421 (1998; Figure~1(c)), 
Mrk~501 (1998; Figure~1(d)), and PKS 2155$-$304 (1996; Figure~1(g)), 
successive flares are clearly seen. 

Figure~2 shows the long-term variation of fluxes, with both
$continuous$ and $short$ \asca and \rxte observations plotted.
\asca observations of Mrk~421 spanned more than 
5~years (from 1993 to 1998), and show that the source exhibits variability by 
more than an order of magnitude. Blow-ups of the light curves taken in 
1995 and 1997 are given in the lower panel (Figure~2 [A],[B]). 
For Mrk~501 and PKS~2155$-$304,  we plot the \rxte data because the 
observations were conducted much more frequently than the \asca observations. 
The \rxte observations spanned 
more than 3~years and significant flux changes are clearly detected.
Blow-ups of light curves for $short$ observations are given in 
Figure~2 [C]$-$[H]. 

\subsection{Power Spectrum Density}

Power Spectrum Density (PSD) analysis is the most common technique used
to characterize the variability of the system. The high quality data obtained 
with \asca and \rxte enable us to determine the PSD over 
a wider frequency range than attempted previously.  
An important issue is the data gaps, which are 
unavoidable for low-orbit X-ray satellites (see Figure~1). Since the 
orbital period of \asca and \rxte is 5760~sec, Earth occultation makes 
periodic gaps every 5760~sec, even for the $continuous$ \asca observations 
(Table~1).  Similarly, the long-look \rxte observations of Mrk~501 and 
PKS~2155$-$304 (Figure~1 (d), (g)) have artificial gaps, since the 
observations 
are spaced typically 3 or 4 orbits (17280 or 23040 sec) apart.  
To reduce the effects caused by such windowing, we introduce a technique 
for calculating the PSD of unevenly sampled light curves. 

Following  Hayashida et al.\ (1998), the NPSD ($Normalized$ Power Spectrum 
Density) at frequency $f$ is defined as
\begin{eqnarray} 
P(f) = \frac{[a^2(f)+b^2(f)-\sigma^2_{\rm stat}/n]T}{F_{\rm av}^2},\nonumber \\
a(f) = \frac{1}{n}\sum_{j=0}^{n-1} F_j {\rm cos} (2\pi f t_j), \nonumber \\
b(f) = \frac{1}{n}\sum_{j=0}^{n-1} F_j {\rm sin} (2\pi f t_j), \nonumber \\
\end{eqnarray}
where $F_j$ is the source count rate at time $t_j$ (0$\le$$j$$\le$$n$$-$1), 
$T$ is the data length of the time series and  $F_{\rm av}$ is the mean 
value of the source counting rate. 
The power due to the photon counting statistics is given 
by $\sigma_{\rm stat}^2$.
With our definition, integration of power over the positive frequencies is 
equal to half of the light curve excess variance (e.g., Nandra et al.\ 1997).

To calculate the NPSD of our data sets, we made light curves of two 
different bin sizes for the \asca data (256 and 5760~sec), and three 
different bin sizes for the \rxte data (256, 5760, and 23040~sec). 
We then divided each light curve into ``segments'', which 
are defined as the continuous part of the light curve.  If the light curve 
contains a time gap larger than twice the data bin size, we cut
the light curve into two segments before/after the gap. 
We then calculate the power at frequencies $f$ = $k$/$T$ 
(1 $\le$ $k$ $\le$ $n$/2) for each segment and take the average. 

In this manner, the light curve binned at 256~sec is divided into 
different segments every 5760~sec, corresponding to the gap due to 
orbital period.  On the other hand, the light curve binned at 5760 
(or 23040) sec is smoothly connected up to the total observation length $T$, 
if further artificial gaps are not involved. 
This technique 
produces a large blank in the NPSD at around 2$\times$10$^{-4}$~Hz 
(the inverse of the orbital period), 
but the effects caused by the sampling window are minimized. 
The validity of the NPSD value at other frequencies are 
discussed in detail in Hayashida et al.\ (1998). 
In the following, we calculate 
the NPSD using the data from the $continuous$ observations in Table~1. 
 
Figure~3 (a)$-$(g) shows the NPSD calculated with this procedure.
The upper frequency limit is the Nyquist frequency (2$\times$10$^{-3}$~Hz 
for 256~sec bins) and the lower frequency is about half the inverse of the 
longest continuous segments. These NPSD are binned in logarithmic intervals 
of 0.2 (i.e.\ factors of 1.6) to reduce the noise. 
The error bars represent the 
standard deviation of the average power in each rebinned frequency interval.
The expected noise power due to counting statistics, 
$\sigma^2_{\rm stat}$$T$/($n$$F_{\rm av}^2$) (see Equation (1)), are shown 
in each panel as a dashed
line. For the \asca data, we calculate the NPSD 
using both the GIS and the SIS data, while the PCA data were used 
for the \rxte light curves (see Figure~1).
 
One finds that the NPSDs follow a power-law that decreases with increasing 
frequency in the high-frequency range. For the long-look observations 
of Mrk~421 (1998; Figure~3(c)), Mrk~501 (1998; Figure~3(d)), and 
PKS~2155$-$304 (1996; Figure~3(g)), 
signs of a roll-over can be seen at the low-frequency end
($f$ $\sim$ 10$^{-5}$~Hz). Since all the NPSDs have very steep power-law 
slopes, 
only little power exists above 10$^{-3}$ Hz. This is very different from 
the PSDs of Seyfert galaxies, 
for which powers are well above the counting noise 
up to 10$^{-2}$~Hz (e.g., Hayashida et al.\ 1998; Nowak \& Chiang 2000). 
Note that this is $not$ due to low counting statistics because the TeV 
sources discussed here are much brighter in X-rays than most Seyfert galaxies. 

To quantify the slope of the NPSD, we first fit a $single$ power-law to each 
NPSD in the frequency range $f$ $\le$ 10$^{-3}$~Hz. We do not use the data 
above 10$^{-3}$~Hz because they tend to be noisy and often consistent with 
zero power. The results are summarized in Table~3. A single power-law function 
turned out to be a good representation of all observations except for 
those of Mrk~421 (1998; Figure~3(c)), Mrk~501 (1998; Figure~3(d)), and 
PKS~2155$-$304 (1996; Figure~3(g)). 
The best fit power-law slopes ($\alpha$ of $f^{-\alpha}$) 
range from $\sim$2 to 3, indicating a strong red-noise behavior. 
For the three long-look observations, this model did not represent 
the NPSD adequately; the power-law fitting the data below 10$^{-5}$~Hz was 
too flat for the data above 10$^{-5}$~Hz. 
For these observations, the $\chi^2$ was 39.7 (23~dof) for 
Mrk~421, 32.2 (11 dof) for Mrk~501, and 22.6 (11 dof) for PKS~2155$-$304. 
A single power-law model is thus rejected with higher than 98$\%$ 
confidence level for these long-look observations.

A better fit was obtained using a $broken$ power-law model, where the spectrum
is harder below the break. The fitting function used was $P(f)$ $\propto$ 
$f^{-\alpha_{\rm L}}$ for $f$ $\le$ $f_{\rm br}$, and $P(f)$ $\propto$ 
$f^{-\alpha}$ for $f$ $>$ $f_{\rm br}$. With this relatively simple model,
the exact shape of the turn-over is not well constrained and the low-frequency 
behavior is undetermined. We thus fixed $\alpha$ at the best fit 
value determined from a power-law fit in the high-frequency range of 
10$^{-5}$ to 10$^{-3}$~Hz, and kept $\alpha_{\rm L}$ and $f_{\rm br}$ as free 
parameters. The fitting results are given in Table~4.  
The goodness of the fit was significantly improved, 21.1 (22~dof), 
18.6 (10~dof), and 3.9 (10~dof), for  Mrk~421, Mrk~501, and PKS~2155$-$304, 
respectively. 
For these three sources, the break frequency ranges from 1.0 to 
3.0$\times$10$^{-5}$~Hz, roughly consistent with the apparent 
time-scale of successive flares  seen in Figure~1. 
Below the break, the slope of the NPSD ($\alpha_{\rm L}$) 
is relatively flat, ranging from 0.9 to 1.5.

Finally, we comment on the effects caused by sampling windows. As 
mentioned above, our PSD technique is less affected by the sampling windows, 
because only the continuous parts of the light curve are used for the 
calculation. In fact, this seems to have negligible effects for the \asca 
data, because the interruptions are almost even and the observing efficiency 
is high ($\sim$ 0.5). However, for the \rxte data, sampling effects
may be significant because the observations are conducted less frequently
and the observing efficiency is low  (0.2 or 0.3). 
The most rigorous estimate of this effect would be obtained by simulating
the light curves characterized with a 
certain PSD, filtered by the same window as the actual observation.
The resulting PSDs could then be compared with that we assumed.  
However, such an estimate is only possible when we already know the 
$true$ PSD of the system. 

As an alternative approach, we approximate each data gap by an interpolation 
of actual data, fitted to a linear function. The gaps in light curve are 
thus bridged in a smooth way over total observation length. We note that, 
even if the data are linearly interpolated across the gaps, 
Poisson errors associated with these points remain quite uncertain. 
We therefore calculated the NPSD in the frequency range $f$ $<$ 2 
$\times$ 10$^{-4}$~Hz, where counting errors becomes negligible 
compared to the power due to 
intrinsic source variability (see the dashed lines in Figure~3).
We tested this interpolation method for the data of the three long-look 
observations.
 
We found that the NPSDs calculated from this interpolation method are 
entirely consistent with those given in Table~4 for both Mrk~421
and PKS~2155$-$304.  For Mrk~501, 
$\alpha$ and $\alpha_{\rm L}$ are consistent, but $f_{\rm br}$ is estimated 
to be (1.3$\pm$0.4)$\times$10$^{-5}$~Hz, which is slightly smaller than the 
value in Table~4 (viz.\ (3.0$\pm$0.9)$\times$10$^{-5}$~Hz). Such a difference, 
however, could be due to poor statistics of the NPSD plots (Figure~3 (d)) 
rather than the sampling effects discussed here. In fact, we have to determine 
$f_{\rm br}$ only from several data points around the turn-over. Moreover, 
although we have fitted the NPSD with $\alpha$ fixed to the best fit value 
of 2.92 (see Table~4), a wider range of $f_{\rm br}$ would be acceptable when 
all parameters are allowed to vary.

Also note that such interpolations might introduce large systematics in the 
resulting PSD when the observing efficiency is low. In fact, interpolations 
of the observed data to fill the gaps would produce the $smoothest$ possible 
solution, because it assumes the least variations across the observational 
data. This might affect the resulting PSD slopes 
($\alpha$ and $\alpha_{\rm L}$) as well as the break frequency ($f_{\rm br}$), 
especially for the \rxte observations.  
The exact position of a break is thus unclear, but conservatively, 
we can give a frequency $f_{\rm br}$ $\simeq$ 10$^{-5}$~Hz with an uncertainty 
factor of a few or larger. 
In the next section, we thus consider a wide range for the roll-overs 
($f_{\rm br}$), using the more powerful structure function technique.

\subsection{Structure Function}

In this section we examine the use of 
a numerical technique called the structure function (hereafter, SF). 
The SF can potentially provide information on the 
nature of the physical process causing any observed variability. 
While in theory the 
SF is completely equivalent to traditional Fourier analysis methods 
(e.g., the PSD; $\S$~2.3), it has several significant advantages. Firstly, 
it is much easier to calculate. Secondly, the SF is less affected by gaps 
in the light curves (e.g., Hughes et al.\ 1992).  
The definitions of SFs and their properties are 
given by Simonetti et al.\ (1985). The first order SF is defined as
\bq0
{\rm SF}(\tau) = \frac{1}{N}\sum[a(t) - a(t+\tau)]^2, 
\label{equation:1-1}
\eq
where $a(t)$ is a point of the time series (light curves) $\{$$a$$\}$ and 
the summation is made over all pairs separated in time by $\tau$. 
$N$ is the number of such pairs. Note that the SF is free from the DC 
component in the time series, whereas techniques such as the auto-correlation 
function (ACF) and the PSD are not.

The SF is closely related with the power spectrum density (PSD) distribution. 
If the structure function has a power-law form, SF($\tau$) $\propto$ 
$\tau^{\beta}$ ($\beta$ $>$ 0), then the  power spectrum has the distribution 
$P(f)$ $\propto$ $f^{-\alpha}$, where $f$ is frequency and $\alpha$ 
$\simeq$ $\beta$ + 1.  We note that this approximation is invalid 
when $\alpha$ is smaller than 1. In fact, both the SF and the NPSD should 
have zero slope for white noise, because it has zero correlation time-scale. 
However, the relation holds within an error of $\Delta\alpha$ $\simeq$ 
0.2 when $\alpha$ is larger than $\sim$1.5 (e.g., Paltani et al.\ 1997; 
Cagnoni, Papadakis \& Fruscione 2000; Iyomoto \& Makishima 2000). 
Therefore the SF gives a crude but convenient estimate of the corresponding 
PSD distribution which characterizes the variability. 

In general, the SF gradually changes its slope ($\beta$) with time interval 
$\tau$. On the shortest time-scale, variability can be well approximated 
by a linear function of time; $a(t)$ $\propto$ $t$.
In this time domain, the resulting SF is $\propto$ $\tau^2$, which is the 
steepest portion in the SF curve (see Equation~(2)). For longer time-scales, 
the slope of the SF 
becomes flatter ($\beta$\,$<$\,2) reflecting the physical process operating 
in the system. When $\tau$ exceeds the longest time variability of the 
system, the SF further flattens, with $\beta$\,$\sim$\,0, which is the 
flattest portion in the SF curve (white noise). At this end, the amplitude 
of the SF is equal to twice the variance of the fluctuation.

In Figure~4, the SFs are plotted for the light curves presented in Figure~1.
\asca (GIS) and \rxte (PCA) light curves binned in 1024~sec intervals
are used for the calculation. 
The resulting SFs are normalized by the square of the mean 
fluxes, and are binned at logarithmically equal intervals.
The measurement noise (Poisson errors associated with flux uncertainty) 
is subtracted as twice the square of Poisson errors on the fluxes; 
2 $<$$\delta$$a^2$$>$. The noise level is shown as a dashed line in the 
figures. All the SFs are characterized with a steep increase ($\beta$ $>$ 1) 
in the time region of 10$^{-2}$ $<$ $\tau$/day $<$ 1, roughly consistent 
with the corresponding NPSDs given in Figure~3 
($P(f)$ $\propto$ $f^{-2 \sim -3}$).

The SFs of the long-look observations show a variety of features. 
For example, the SF of Mrk~421 (Figure 4(c)) shows a very complex SF that 
cannot even be described as a simple power-law, as it flattens around 
0.5 day, then steepens again around 2 days. A similar ``roll-over'' can be 
seen for Mrk~501 and PKS 2155$-$304 around 1~day (Figure~4 (d),(g)). 
Importantly, these turn-overs reflect the typical time-scale of 
repeated flares, corresponding to the break in the NPSDs described 
in $\S$~2.3. The complicated features (rapid rise and decay)
at large $\tau$ may not be real and may result from the 
insufficiently long sampling of data. The number of pairs in Equation~(1) 
decreases with increasing $\tau$, and hence the resulting SF becomes uncertain 
as $\tau$ approaches $T$, where $T$ is the length of time series. 
The statistical significance of these features can be tested using the 
Monte Carlo simulation described below.

We next calculate the structure functions using the total light curves 
given in Figure~2. Using 5 year's \asca data and  3 year's \rxte data,
we can investigate the variability in the widest time domain over more than 
five orders; $10^{-2}$ $\le$ $\tau$/day $\le$ 10$^3$.
The results are respectively given for Mrk~421, Mrk~501 and PKS~2155$-$304 
in Figure~5. $Filled$ $circles$ are observational data, 
normalized by the square 
of the mean fluxes, and are binned at logarithmically equal intervals.
All the SFs show a rapid increase up to $\tau$/day $\simeq$ 1, then gradually 
flatten to the observed longest time-scale of $\tau$/day\,$\ge$\,1000. 
Fluctuations at large $\tau$ ($\tau$/day\,$\ge$\,10) are due to the
extremely sparse sampling of data. In fact, even for the case of Mrk~501 
(the most frequently sampled data), the total observation time is 700~ksec, 
which is only 1$\%$ of the total span of 3~years. Although we cannot 
apply the usual PSD technique to such under-sampled data, it appears
the SF still can be a viable estimator.   

In order to demonstrate the uncertainties caused by such sparse sampling, and 
to firmly establish the reality of the ``roll-over'', we simulated 
the long-term light curves (Figure~2) following the $forward$ $method$ 
described in Iyomoto (1999).  We first assume a certain PSD which describes 
the characteristic variability of the system. Using a Monte Carlo technique, 
we generate a set of random numbers uniformly distributed between 0 and 
2$\pi$ and use them as the random phases of the Fourier components. 
A fake light curve is then generated by a Fourier transformation, 
with the constraint that the power in each frequency bin decreases as 
specified by the PSD. We simply choose a deterministic amplitude for each 
frequency and randomize only the phases, a common approach
(e.g., Done et al. 1989). It may be most rigorous to also assume 
``random amplitudes'' distributed within 1~$\sigma$ of the input PSD 
(Timmer \& K\H{o}nig 1995), but simulations based on their algorithm 
remain as a future work.

The resulting light curve is filtered by the same sampling window as the 
actual observation, and is normalized to have same RMS (root mean
square) as the actual data. We repeat this process using
different sets of random numbers and generate 1000 light curves for the 
assumed PSD. Finally the SFs are calculated for the individual light
curves. We found that simulated SFs generally show the kinds of bumps and
wiggles as the real data, and sometimes show a roll-over even if none 
was simulated. Several examples of simulated SFs are shown in Figure 6.
Such ``structures'' often appeared at large $\tau$,
probably due to finite length of the light curve. We perform the same 
statistical test to these simulated data for quantitative comparison with
the actual SF. 

We first applied this technique assuming a PSD of the form $P(f)$ $\propto$ 
$f^{-\alpha}$, where $\alpha$ is determined from the best fit NPSD parameters 
given in Table~4. In order to reproduce the long-term light curves of 
Mrk~421, Mrk~501 and PKS~2155$-$304, we take $\alpha$\,= 2.1, 2.9, and 2.2, 
respectively. Based on a set of a thousand fake light curves, we
computed the expected mean value, $<$$SF_{\rm sim}(\tau)$$>$, and variance, 
$\sigma$$_{SF(\tau)}$, of all the simulated SFs at each $\tau$. 
The results are superimposed in Figure~5 as $crosses$.  
Errors on simulated data points are equal to
$\pm$$\sigma$$_{SF(\tau)}$. One finds that errors become larger at large
$\tau$, meaning that the SF tends to involve fake bumps and wiggles near 
the longest observed time-scale. Also note that we cannot use these errors in 
the normal $\chi^2$ estimation, 
since the actual SF is $not$ normally distributed. Large deviations
between the actual SFs ($filled$ 
$circles$) and the simulated ones ($crosses$) are apparent, but
quantitative comparison with actual data is necessary.

To evaluate the statistical significance of the goodness of fit, and 
to test the reality of complicated features in the SF, we then calculate
the sum of squared differences, 
$\chi_{\rm sim}^2$ = $\sum_{k}$$\{$log[$<$$SF_{\rm sim}$($\tau_{k}$)$>$]$-$log[$SF$($\tau_{k}$)]$\}$$^2$. 
Strictly speaking, ``$\chi_{\rm sim}^2$'' defined here is different from 
the traditional $\chi^2$, but the statistical 
meaning is the same. For the actual SFs, these values are  
$\chi_{\rm sim}^2$ = 1608, 702, and 521 for Mrk~421, Mrk~501 and 
PKS 2155$-$304, respectively. We then generated $another$ set of 1000 
simulated light curves and hence fake SFs to evaluate the the distribution
of $\chi_{\rm sim}^2$ values. From this 
simulation, the probability that the X-ray light curves are the realization 
of the assumed PSDs is $P(\chi^2)$ $<$ 10$^{-3}$ (0 of 1000 simulated light 
curves) -- none of which are good expression of the data. 

We thus introduce a ``break'', below which the slope of the 
PSD becomes flatter. Similar to  $\S$~2.3, we assume a  PSD of the form, 
$P(f)$ $\propto$ $f^{-\alpha_{\rm L}}$ for $f$ $<$ $f_{\rm br}$, and 
$P(f)$ $\propto$ $f^{-\alpha}$ for $f$ $>$ $f_{\rm br}$, where
$\alpha$ and $\alpha_{\rm L}$ were set to the best fit value given 
in Table~4, namely ($\alpha_{\rm L}$, $\alpha$) = (0.9, 2.1) for Mrk~421, 
(1.4, 2.9) for Mrk~501, and (1.5, 2.2) for PKS~2155$-$304, respectively. 
Since the exact position of a break is not well constrained, we simulate 
various cases of $f_{\rm br}$ = 3.9$\times$10$^{-5}$, 1.2$\times$10$^{-5}$, 
and  3.9$\times$10$^{-6}$~Hz, which correspond to the break in the SF at 
$\tau$/day $\simeq$ 0.3, 1, 3, respectively. 

As a result, the statistical significance is significantly improved. 
Results are given in Figure~5 as $open$ $squares$.
For Mrk~421, $\chi^2$ of the actual data is minimized when $f_{\rm br}$ 
= 3.9$\times$10$^{-6}$ Hz ($\chi^2$ = 47; $P(\chi^2)$ = 0.59), but 
other values for $f_{\rm br}$ are not a good representation of data. 
For Mrk~501, both $f_{\rm br}$ = 1.2$\times$10$^{-5}$ and 
3.9$\times$10$^{-6}$ Hz are acceptable in the meaning that 
0.1 $<$ $P(\chi^2)$ $<$ 0.9 ($\chi^2$ = 75 and 71, respectively).
Similarly, the SF of PKS~2155$-$304 is acceptable for the break of 
$f_{\rm br}$ = 1.2$\times$10$^{-5}$ and 3.9$\times$10$^{-6}$ Hz, 
with $\chi^2$ = 21 and 43 ($P(\chi^2)$ = 0.74 and 0.57, respectively). 
To obtain an upper limit of the variability time scale, we further  
tested the case when $f_{\rm br}$ = (0.3$-$1)$\times$10$^{-6}$ Hz, 
which corresponds to the break in the SF at $\tau$/day $\simeq$ 10$-$30;
none of which were turned out to be acceptable. 
Thus, although the exact position of the break is still uncertain, 
the possibility  that the X-ray light curves
are the realizations of a $single$ power-law 
PSD can be rejected.  We thus conclude that (1) the PSD of the TeV sources 
have at least one roll-over at 10$^{-6}$ Hz $\le$ $f_{\rm br}$ $\le$ 
10$^{-5}$ Hz (1 $\le$ $\tau$/day $\le$ 10),  and (2) the PSD changes 
its slope from $\propto$ $f^{-1 \sim -2}$ ($f$ $<$ $f_{\rm br}$) to $\propto$ 
$f^{-2 \sim -3}$ ($f$ $>$ $f_{\rm br}$) around the roll-over.

We finally refer to the long-term variability of Mrk~501 and their sampling 
pattern (Figure~2). As mentioned in $\S$~2.1, Mrk~501 was in a
historically high state in 1997, with the result that three of the four 
observations 
listed in Table~2 are (more or less) intentionally conducted during this high 
state. The SF takes existing data points at certain epochs out of an (unknown) 
$true$ variation, implicitly assuming them to be representative of the system.
Strictly speaking, the SF analysis will only be valid if the epochs of the 
observations are randomly chosen regardless of activity of the system.  This 
seems not to be the case with the \rxte data of Mrk~501 because observations 
are $biased$ to the high state in 1997 (125~ksec of the total 700~ksec exposure). 
To see the effects caused by this sampling pattern, we also performed the SF  
analysis only using the data taken in 1998. The resulting SF had a very
similar shape, but the absolute value of the SF at each $\tau$ tended to be 
smaller by a factor of $\sim$ 5, which means that amplitude of variation is 
smaller by a factor of $\sim$ 2  in 1998 (see Figure~2).
In spite of such difference, the most important part of the result did not 
change: the PSD of Mrk~501 needs at least one roll-over at 
10$^{-6}$ Hz $\le$ $f_{\rm br}$ $\le$ 10$^{-5}$ Hz.

\section{Discussion}

\subsection{Comparison with Previous Works}

As seen in $\S$~2, the short time-scale variability of the TeV sources 
can be described by a steep PSD index up to the characteristic
time-scale of order or longer than 1 day. 
This is evidence that only little variability exists on time-scales shorter 
than $t_{\rm var}$, indicating strong red-noise type behavior. 
In this section, we compare our results with those given 
in the literature. 

PKS~2155$-$304 is the only TeV blazar for which PSD studies had previously
been made 
in the X-ray band. Tagliaferri et al.\ (1991) analyzed $EXOSAT$ 
data (exposure of $\sim$1~day) and found that the power spectrum follows 
a power-law with an index $-$2.5$\pm$0.2. Hayashida et al.\ (1998) derived 
the PSD of PKS~2155$-$304 from a \ginga observation. They reported a 
PSD index of $-$2.83$^{+0.35}_{-0.24}$. In the optical, Paltani et al.\ (1997) 
studied the variability based on 15 nights data. They found that the PSD of 
optical data follows a steep power-law of an index $-$2.4 as well. 
In summary, PSDs in the literature showed featureless red-noise
spectra, which are comparable with our results. Very recently, Zhang et al.\ 
(1999) analyzed three X-ray light curves obtained with \asca and \bepposax.
They reported a steep PSD index of $-$2.2 for two data sets, but an 
exception was found during the \bepposax observation in 1997, which yielded
a relatively flat PSD slope of $-$1.54$\pm$0.07. 

For Mrk~421 and Mrk~501, no PSD studies have been reported 
for X-ray variability prior to our work. In the extreme UV band, 
Cagnoni, Papadakis and Fruscione (2000) analyzed Mrk~421 data obtained 
with the $Extreme$ $Ultraviolet$ $Explorer$ over a four-year period.
They reported 
that the PSD of Mrk~421 is well represented by a slope of $-$2.14$\pm$0.28 
with a break at $\sim$3 day.  Similarity of the characteristic time-scale 
and PSD slopes both in the X-ray and EUV bands is very interesting, 
because it may prove that the emission site of the X-ray and EUV
photons are same (or very close) in the relativistic jet (see $\S$~3.3).
Although high under-sampling does not allow us to test whether or not
the break frequency is the same in both bands, future studies 
based on larger sample of data could clarify this point. 

Also, it is interesting to note that a break in the SFs
and PSDs occurs on the same time-scale (of about 20$-$30 hour) 
is seen in other blazars in different energy regimes.  The frequent 
occurrence of this preferred time-scale in the optical regime 
(Wagner \& Witzel 1995; Heidt \& Wagner 1996) is particularly noteworthy, 
since optical observations are clearly intrinsic and not affected by
interstellar scattering.  They are almost certainly unaffected from 
contributions of inverse-Compton scattering as 
well, and the break in the SF occurs at the same 
time-scale as observed here for sources in which the synchrotron component
extends well into the X-ray regime. This suggests that the time-scale
is not due to $\gamma$-$\gamma$ pair absorption or any process which is
correlated to the cut-off frequency of the synchrotron branch. The
phenomenon of Intra-Day Variability (indicating maximum amplitudes for
variations on time-scales of the order of 1~day; Wagner \& Witzel 1995) 
clearly extends into the X-ray regime as well.

Finally, we briefly comment on the red-noise leak, which might be 
important to characterize the variability in blazars. In general, the PSD 
is biased estimator due to the finite length of real observations. In 
particular, it is pointed out that large amounts of power could leak through 
from low to high frequencies when $\alpha$ of $f^{-\alpha}$ is larger than 2 
(strong red noise; Deeter \& Boynton 1982).  For example, Papadakis 
\& Lawrence (1995) studied a simulated 
time series with a power spectrum that followed $f^{-2.8}$ and flattened
below a certain frequency. They found that when the data length is shorter 
than 10 times the break time-scale (which is very similar to our situation), 
the resulting PSD is biased. A considerable amount of power has 
been transferred from low to high frequencies, and the resulting PSD follows 
a flatter slope with $\alpha$~=~2.4. In order that resulting PSD not be 
biased, the observational data length must be longer than 100 times the break 
time-scale. 
 
These studies suggest that the NPSDs derived in this paper may
also be biased, and the actual slopes may be $steeper$ than we have 
estimated.  In fact, the steep PSD could be the result of ``red-noise'' leak 
even if variability shorter than the characteristic time-scale is really 
absent. On the other hand, it might be explained by the superposition of 
rapid  $micro$ $flares$ (e.g., less than 10$\%$ fluctuations) 
which are hidden behind large flares occuring on a time-scale of the
order of 1 day or longer (factor of $\sim$~2). 
At present, we cannot discriminate between 
these situations. Time series analysis with much better photon 
statistics, as well as more detailed simulations would clarify this point.
Work along these lines is now in progress (Tanihata et al.\ 2000, 
in preparation), but is beyond the scope of this paper.  

\subsection{Seyferts versus Blazars}

Comparing our results with those of other black-hole systems is quite 
interesting, as it is well known that the PSDs of Seyfert galaxies and 
Galactic black-holes in the X-ray band also behave as power-laws over some 
temporal frequency range.  Our results (Table~3 and 4) 
are contrasted with those of Hayashida et al.\ (1998) in Figure~7.
We note that Hayashida et al.\ (1998) fit the NPSD with a single 
power-law in the frequency range $f$ $\ge$ 10$^{-5}$ Hz. Lower frequency 
data were not available because a typical \ginga observation lasted only 
one-day with longer data-sets containing large gaps.  
The situation is similar for
the typical \asca and \rxte observations, but not for the three long-look 
observations.  
These data lower the frequency limit to 10$^{-6}$\,Hz 
(Figure 3 (c),(d),(g)). To make a quantitative comparison with Hayashida
et al.\ (1998), we thus measured the PSD slope from a power-law fit in the 
frequency range $f$ $\ge$ 10$^{-5}$ Hz; these are  
2.14$\pm$0.06, 2.92$\pm$0.27 and 2.23$\pm$0.10, respectively for the
case of Mrk 421, Mrk 501 and 
PKS~2155$-$304 ($\alpha$ of Table~4). 
  
We find that the PSD slopes of the TeV sources are clearly different from 
those of Seyfert galaxies and Galactic black-holes, on time-scales shorter
than 1~day. 
Quasi-fractal behavior (1 $<$ $\alpha$ $<$ 2) is a 
general characteristic of Seyfert galaxies and Galactic black-holes, 
while the power-law indices are steeper for the TeV emitting sources 
(2 $<$ $\alpha$ $<$ 3).  This presumably reflects the different physical 
origins of and/or locations for the X-ray production.
In fact, Seyfert galaxies and Galactic black-holes are believed to emit
X-ray photons nearly isotropically from the innermost parts of the accretion 
disk  (see e.g., Tanaka et al.\ 1995; Dotani et al.\ 1997), 
while non-thermal emission from a relativistically beamed jet 
is the most likely origin of X-rays for blazar-like sources (see $\S$~1).  

We note that Hayashida et al.\ (1998) have estimated black hole masses 
in various types of AGNs using time variability.  A linear proportionality 
between the variability time-scale and the black hole mass was assumed, 
and this relation for Cyg~X-1 ($M$ $\simeq$ 10 $M_\odot$) was used as a 
reference point. Such an approach may be viable for AGNs for which the 
emission mechanisms are thought to be similar to Galactic black hole systems 
(e.g., Seyfert Galaxies), but not for the blazar class. 
Indeed, the masses derived by their method for the blazars 3C~273 and 
PKS~2155$-$304 indicated that the observed 
bolometric luminosity exceeds the Eddington limit; this can be interpreted 
as indicating the importance of beaming effects in these objects. 
To estimate the mass of the central engine postulated to exist in blazars, 
a completely different approach must be applied, as discussed below.

\subsection{Implication for the Mass of Central Engine}

As first pointed out by Kataoka (2000), the little power of rapid 
variability ($\le$~1~day) in TeV sources provides important clues to the 
X-ray emitting site in the jet. The characteristic time-scale of each 
flare event $t_{\rm var}$ $\ge$ 1 day should reflect the size of the emission 
region (see below), which we infer to be $\ge$ 10$^{16}$ ($\Gamma$/10)~cm in 
the source co-moving frame, if emitting blobs are approaching with Lorentz 
factors $\Gamma$ ($\Gamma$ $\simeq$ 10; Vermeulen \& Cohen 1994). 
This range of Lorentz factors is independently inferred from 
constraints that can be derived from the spectral shape and variability of 
TeV blazars (e.g., Tavecchio, Maraschi and Ghisellini 1998; Kataoka et al.\ 
1999). 

If the jet is collimated to within a cone of constant opening angle 
$\theta$ $\simeq$ 1/$\Gamma$ and the line-of-sight extent of shock 
is comparable with the angular extent of the jet,  one expects that the 
X-ray emission site is located at distances $D$ $\ge$ 10$^{17}$ 
($\Gamma$/10)$^{2}$ cm from the base of the jet. 
Only little variability shorter than $t_{\rm var}$ 
strongly suggests that no significant X-ray emission can occur in regions 
closer than this to the black hole. The relativistic electrons responsible 
for the X-ray emission are most likely accelerated and injected at shock 
fronts occurring in the jet (e.g., Inoue \& Takahara 1996; Kirk, Rieger \& 
Mastichiadis 1998; Kusunose, Takahara \& Li 2000). The lack of short term 
variability may then imply that shocks are nearly absent until distances of 
$D$ $\ge$ 10$^{17}$ ($\Gamma$/10)$^{2}$ cm. Two different ideas have 
been put forward as to how and where shocks form and develop in blazar jets:
external shocks and internal shocks. Both have also been extensively discussed 
in relation to gamma-ray bursts.

Dermer \& Chiang (1998) (see also Dermer 1999) have proposed an $external$ 
$shock$ model, wherein the shocks arise when outflowing jet plasma
decelerates upon interaction with dense gas clouds 
originating outside the jet. The precise nature of the required gas clouds 
is uncertain, but they may be similar to the ones 
postulated to emit the broad emission lines in Seyfert galaxies and quasars.
It is interesting to note that the location of the broad line regions 
in such strong emission line objects have been inferred to be 10$^{17-18}$ cm 
from the nucleus (e.g.,  Ulrich, Maraschi \& Urry 1997), in line with the 
distances presented above. It remains to be seen whether this picture is 
viable for the BL~Lac objects considered here;
gas clouds may be more dilute or absent in such objects 
as suggested by the weakness of their emission lines
(e.g. B\"ottcher \& Dermer 1998 and references therein).
However, this could instead be due to a weaker central ionizing source
rather than a difference in gas cloud properties.

An alternative view concerns $internal$ $shocks$, originally invoked to 
explain the optical knots in the jet of M87 (Rees 1978). Ghisellini (1999; 
2000) has pointed out that this idea successfully explains some observed 
properties of blazars. In this scenario, it is assumed that the central 
engine of an AGN intermittently expels blobs of material with varying bulk 
Lorentz factors rather than operating in a stationary manner.

Consider, for simplicity,  two relativistic blobs 
with bulk Lorentz factors $\Gamma$ and $a_0$$\Gamma$ ($a_0$ $>$ 1)
ejected at times $t$ = 0 and $t$ = $\tau_0$ $>$ 0, respectively.
The second, faster blob will eventually catch up and collide 
with the first, slower blob, leading to shock formation and generation of a 
corresponding X-ray (and possibly TeV) flare
(A more realistic situation would envision sequential ejections of many blobs 
inducing multiple collisions and a series of flares; e.g., Figure 1 (c)).
The time interval between the two ejections is determined by
the variability of the central engine and is expected to be 
at least of the order of the dynamical time close to the black hole, 
i.e. approximately the Schwarzschild radius ($R_{\rm g}$) light crossing 
time. Writing $\tau_0$ = $k$ $R_{\rm g}$/c where $k$\,$\ge$\,3, 
the distance $D$ at which the two blobs collide is

\bq0 
D = c \tau_0 \Gamma^2 (\frac{2 a_0^2}{a_0^2 - 1}) 
  = 10^3 (\frac{k}{10}) (\frac{\Gamma}{10})^2 (\frac{2 a_0^2}{a_0^2 - 1})
R_{\rm g}.
\label{equation:10-6}
\eq
The radius of the jet at $D$ is 
\bq0
R = D {\rm sin} \theta \simeq D \theta \simeq D/\Gamma, 
\label{equation:10-7}
\eq
which is taken to be equal to the emission blob size.
Accounting for time shortening by a factor $\simeq$ 1/$\Gamma$ due to beaming, 
the observed variability time-scale should be 
\bq0
t_{\rm var} \simeq \frac{D}{c \Gamma^2}
  =    10 (\frac{k}{10}) (\frac{2 a_0^2}{a_0^2 - 1}) \frac{R_{\rm g}}{c}.
\label{equation:10-8}
\eq

Substantial dissipation and radiation of the jet kinetic energy 
will not take place until the blobs collide, 
and this can only occur above a minimum distance $D$, 
delimited by the minimum value of $k$. 
It is then a natural consequence that 
variability on time-scales shorter than a certain value 
($D$/($c$$\Gamma^2$))
is suppressed, as indicated from the temporal studies presented in this paper. 

It is apparent in Equation (5) that the minimum variability time-scale depends
on the mass of the central black hole. Taking the typical observed value, 
we obtain 
\bq0
M \simeq 9 \times 10^8 (\frac{t_{\rm var}}{{\rm day}}) (\frac{10}{k}) 
(\frac{a_0^2 - 1}{2 a_0^2}) M_\odot.
\label{equation:10-13}
\eq
In Figure~8, we plot the black hole mass $M$ as a function of $a_0$ for
various parameter sets $k$ and $t_{\rm var}$. 
Even when assuming a wide range of parameters 
($k$ = 5, 20, 100 and $t_{\rm var}$/day = 1, 10), 
the mass of the central black hole is well constrained to 
10$^7$ $<$ $M$/$M_\odot$ $<$ 10$^{10}$.  

The discussion presented above is based on the assumption
that the characteristic time-scale of X-ray variability is indicative of 
the size of the emission region $t_{\rm crs}$ (= $R/c$). On the other hand, 
one might imagine that this time-scale also reflects the electron synchrotron 
cooling time ($t_{\rm cool}$), the electron acceleration time-scale in the 
shock ($t_{\rm acc}$) and/or the time-scale over which the accelerated 
electrons are injected into the emission region. 
In general terms, this is true, but we believe that relaxation 
of local variability by light-travel time effects inside the emitting blob
must be the dominant effect, particularly in the X-ray band.
 
For example, Kataoka et al.\ (2000) discovered that the duration of a flare 
observed in~PKS 2155$-$304 (Figure~1(f)) is the $same$ in different 
X-ray energy bands, which is at odds with a picture in which the rise time 
and decay time of the flare are $directly$ associated with $t_{\rm acc}$ and 
$t_{\rm cool}$ respectively, both of which should be energy-dependent. 
Furthermore, they found that $t_{\rm cool}$ ($\simeq$ $t_{\rm acc}$ at the 
highest electron energy) for the X-ray emitting electrons is much shorter 
than $t_{\rm crs}$, resulting in the quasi-symmetric flare light curves 
often observed in these TeV sources (e.g., Figure 1 (c),(f)).
Taking into account the fact that rapid variability faster than $t_{\rm var}$ 
is suppressed (Figure~3), the $t_{\rm var}$ of 
TeV blazars is most probably characterized by $R/c$.  

We also note that extremely rapid variability on sub-hour time-scales 
has sometimes been observed for Mrk~421 and Mrk~501 
in TeV gamma-rays (Gaidos et al.\ 1996; Quinn et al.\ 1999), 
and more recently in X-rays as well (Catanese \& Sambruna 2000). 
These types of events, which may be relatively rare,
could perhaps be interpreted as shocks forming 
with strongly anisotropic geometries, albeit with a low duty cycle.
If the line-of-sight extent of the shock $d$ was much thinner 
than the angular extent of the jet $R$, 
the observed time-scale could be $\sim$ $d/c$ rather than $R/c$ 
(e.g., Salvati, Spada \& Pacini 1998).  

At the opposite end of the spectrum, our temporal studies show that the SFs 
of TeV blazars continue to $increase$ with flatter slopes on longer 
time-scales ($t$ $\ge$ $1/f_{\rm br}$; Figure 5). 
This means that in addition to the day time-scale flares, 
the TeV sources manifest variations in the baseline 
component. Such long-term variability may possibly be associated 
with that occurring in the accretion disk, e.g., various instabilities 
of the disk, or changes in the accretion rate.  As the putative launching site 
of the jet, the accretion disk should inevitably exert a strong influence
and could significantly modulate the process of jet plasma ejection.
It is noteworthy that the time profiles of TeV blazars become
more similar to those of Seyfert galaxies and Galactic black holes 
for time-scales longer than $\sim$1~day.

\section{Conclusions}

We have studied the X-ray variability of three TeV $\gamma$-ray sources, 
Mrk~421, Mrk~501, and~PKS 2155$-$304, in the widest 
time domain possible, from $10^{3}$ to $10^{8}$~sec. 
Our analyses show clear evidence for a ``roll-over'' with a time-scale 
of order or longer than 1~day, both in the power spectra and the
structure functions.  
Importantly, these ``roll-overs'' can be interpreted as the characteristic 
time-scale of successive flare events. We discovered that below this 
time-scale, there is only small power in the variability, as indicated 
by steep PSDs of $f^{-2 \sim -3}$. 
This is very different from other types of mass-accreting systems 
for which the variabilities are well represented by a fractal 
nature. Our results suggest that the X-rays are $not$ generated throughout 
the jet, but are $preferentially$ radiated from distances of $D$ $\ge$ 
10$^{17}$~cm from the jet base, if emission blobs have bulk Lorentz 
factors $\Gamma$\,$\simeq$\,10.  As a possible interpretation of the 
variability in TeV blazars, the internal shock scenario was discussed. The 
observational properties can be consistently explained if the masses of 
the central black holes are $M$ $\simeq$ 10$^{7-10}$ $M_\odot$, and the 
shocks start to develop at $D$ $\ge$ 10$^{2}$$R_{\rm g}$. 
Similar temporal studies at different wavelengths, 
from radio to $\gamma$ rays, as well as for different classes 
of blazars, will be valuable to discriminate between various emission 
models for blazars, as well as to provide important clues to the dynamics 
of jets. 

\acknowledgments

We greatly appreciate an anonymous referee for his/her helpful comments 
and suggestions to improve the manuscript. 
J.K.\ acknowledges a Fellowship of the Japan Society for 
Promotion of Science for Japanese Young Scientists.

\clearpage

\figcaption[]{X-ray flux variations of TeV blazars in different observations:
(a) Mrk~421 (1993 May 10--11:\asca ),  (b) Mrk~421 (1994 May 16--17:\asca), 
(c) Mrk~421 (1998 Apr 23--30:\asca),   (d) Mrk~501 (1998 May 15--29:\rxte),  
(e) PKS~2155$-$304 (1993 May 3--4:\asca), 
(f) PKS~2155$-$304 (1994 May 19--21:\asca), and 
(g) PKS~2155$-$304 (1996 May 16--28:\rxte).
Observation logs are given in Table~1. 
For the \asca data, the energy range is 0.7$-$10~keV, the count rates from 
both GIS detectors are summed, and the data are binned in 256~sec intervals.
For the \rxte data, the energy range is 2.5$-$20~keV, the count rates from 
3 PCUs are summed, and the data are binned in 5760~sec intervals.
\label{fig1}}

\figcaption[]{Long-term flux variation of three TeV sources.
Mrk~421        (1993 May 10 $-$ 1998 Apr 30 with \asca),
Mrk~501        (1996 Aug 1  $-$ 1998 Oct 2  with \rxte), and  
PKS~2155$-$304 (1996 May 16 $-$ 1998 Jan 13 with \rxte). 
Energy ranges are the same, and count rates were determined the same way,
as for Figure~1.
Observation logs for each  
parentheses are  given in Table~1, while logs for square brackets are given 
in Table~2. Small lower panels are expanded plots of the light curves. 
\label{fig2}}

\figcaption[]{Normalized PSD calculated from the light 
curves in Figure~1.  For the \asca data, both GIS and SIS data are used 
for the calculation.
(a) Mrk~421 (1993 May 10--11:\asca ),  (b) Mrk~421 (1994 May 16--17:\asca),
(c) Mrk~421 (1998 Apr 23--30:\asca),   (d) Mrk~501 (1998 May 15--29:\rxte),
(e) PKS~2155$-$304 (1993 May 3--4:\asca),
(f) PKS~2155$-$304 (1994 May 19--21:\asca), and 
(g) PKS~2155$-$304 (1996 May 16--28:\rxte). 
Measurement noise, at the level shown by the dashed line in each figure,
has been subtracted from each point. 
The best fit power-law function or broken power-law is given as dotted 
lines.\label{fig3}}

\figcaption[]{Structure functions calculated from the light 
curves in Figure~1. Each SF is normalized by the square of the mean fluxes.
(a) Mrk~421 (1993 May 10--11:\asca ),  (b) Mrk~421 (1994 May 16--17:\asca),
(c) Mrk~421 (1998 Apr 23--30:\asca),   (d) Mrk~501 (1998 May 15--29:\rxte),
(e) PKS~2155$-$304 (1993 May 3--4:\asca),
(f) PKS~2155$-$304 (1994 May 19--21:\asca), and 
(g) PKS~2155$-$304 (1996 May 16--28:\rxte).
Measurement noise, at the level shown by the dashed line in each figure,
has been subtracted from each point. 
\label{fig4}}

\figcaption[]{Structure functions of Mrk~421, Mrk~501, and PKS~2155$-$304,  
based on long-term light curves presented in Figure~2. 
$Filled$ $circles$ represent the observational data; 
$crosses$ represent  simulated SFs assuming a single power-law 
NPSD (1/$f_{\rm br}$ = $\infty$); and $open$ $squares$
represent simulated SFs assuming a broken power-law NPSD 
(1/$f_{\rm br}$ = 1 or 3 day).  
Each SF is normalized by the square of the mean fluxes.
Measurement noise, at the level shown by the dashed line in each figure,
has been subtracted from each point. 
\label{fig5}}

\figcaption[]{Examples of $simulated$ SFs for Mrk 421 data taken in 
1993$-$1998. The corresponding $observational$ results are shown 
in Figure 5 (Mrk 421). The PSD is assumed to have a broken power-law where 
the break time scale is 3 day. Full details are given in the text.
\label{fig6}}

\figcaption[]{Comparison of PSD slopes for various black-hole systems. 
$Open$ $circles$ : Hayashida et al.\ (1998). 
$Filled$ $circles$ : PSD slope $\alpha$ determined from a $single$ 
power-law fit (Table 3 and 4 ; this work).  
To make quantitative comparison with Hayashida et al. (1998), we 
limit the fitting range to 10$^{-5}$Hz $\le$ $f$ $\le$ 10$^{-3}$ Hz. 
Full  details are given in the text. 
$Crosses$ : PSD slope $\alpha_{\rm L}$ in the $low$-frequency range 
determined from a $broken$ power-law fit (Table~4 ; this work). 
\label{fig7}}

\figcaption[]{The black hole mass $M$ plotted as a 
function of $a_0$, where $a_0$ is the ratio of velocity of 
blob-1 and blob-2 (for more details, see the text). The mass was estimated for 
various parameter sets; $k$ = 5, 20, 100  and $t_{\rm var}$/day = 1, 10.
\label{fig8}}


\begin{table}
\begin{center}
\caption{Observation log ($continuous$ observations) \label{tbl-1}}
\begin{tabular}{rrrrl}
\tableline\tableline
            &          &                               & Exposure$^a$ &     \\
Source      &Satellite & Observing Time (UT)           &(ksec)&Figure   \\
\tableline

Mrk~421        & \asca &1993.05.10 03:22$-$05.11 03:17 &  43  &  1(a)\\
               & \asca &1994.05.16 10:04$-$05.17 08:06 &  39  &  1(b)\\
               & \asca &1998.04.23 23:08$-$04.30 19:32 &  280 &  1(c)\\
Mrk~501        & \rxte &1998.05.15 12:34$-$05.29 12:11 &  306 &  1(d)\\
PKS~2155$-$304 & \asca &1993.05.03 20:56$-$05.04 23:54 &  37  &  1(e)\\
               & \asca &1994.05.19 04:38$-$05.21 07:56 &  96  &  1(f)\\
               & \rxte &1996.05.16 00:40$-$05.28 15:26 &  161 &  1(g)\\
\tableline
\end{tabular}
\tablenotetext{a}{Exposure of GIS for \asca and PCA for \rxte.}
\end{center}
\end{table}

\begin{table}
\begin{center}
\caption{Observation log ($short$ observations) \label{tbl-1}}
\begin{tabular}{rrrrl}
\tableline\tableline
            &          &                               & Exposure$^a$ &  \\
Source      &Satellite & Observing Time (UT)           &(ksec)& Figure  \\
\tableline
Mrk~421        & \asca &1995.04.25 19:16$-$05.08 13:27 &  91  &  2[A]\\
               & \asca &1997.04.29 01:45$-$05.06 08:32 &  70  &  2[B]\\
Mrk~501        & \rxte &1997.04.03 04:27$-$04.16 10:51 &  36  &  2[C]\\
               & \rxte &1997.05.02 04:19$-$05.15 06:49 &  51  &  2[D]\\
               & \rxte &1997.07.11 23:23$-$07.16 04:55 &  38  &  2[E]\\
               & \rxte &1998.02.25 17:29$-$10.02 23:06 & 262  &  2[F]\\
PKS~2155$-$304 & \rxte &1996.11.14 07:39$-$11.24 13:12 &  74  &  2[G]\\
               & \rxte &1998.01.09 03:07$-$01.13 14:46 &  11  &  2[H]\\
\tableline
\end{tabular}
\tablenotetext{a}{Exposure of GIS for \asca and PCA for \rxte.}
\end{center}
\end{table}

\begin{table}
\caption{Fit results of the NPSD with a $single$ power-law\label{tbl-2}}
\begin{center}
\begin{tabular}{rccr}
\tableline\tableline
Source      & Observation$^a$    & $\alpha$$^b$& $\chi^2$ (dof) \\
\tableline
Mrk~421        & \asca 1993 &   2.56$\pm$0.09 & 18.2 (11) \\
               & \asca 1994 &   2.14$\pm$0.24 & 15.2 (10) \\
               & \asca 1998 &   2.03$\pm$0.03 & 39.7 (23)$\dagger$ \\       
Mrk~501        & \rxte 1998 &   1.88$\pm$0.07 & 32.2 (11)$\dagger$ \\
PKS 2155$-$304 & \asca 1993 &   2.14$\pm$0.22 & 3.9 (4)   \\
               & \asca 1994 &   3.10$\pm$0.20 & 17.2 (15) \\
               & \rxte 1996 &   1.90$\pm$0.03 & 22.6 (11)$\dagger$ \\
\tableline
\end{tabular}
\tablenotetext{a}{GIS data (0.7--10 keV) and SIS data (0.5--10 keV) were 
used for \asca observations and PCA data (2.5--20 keV) were used for 
\rxte observations.}
\tablenotetext{b}{The best fit power-low index of NPSDs ($\alpha$ of 
$f^{-\alpha}$).}
\tablenotetext{\dagger}{Goodness of the fit is bad, with $P(\chi^2)$ $<$ 0.02.}
\end{center}
\end{table}

\begin{table}
\caption{Fit results of the NPSD with a $broken$ power-law \label{tbl-2}}
\begin{center}
\begin{tabular}{rccccr}
\tableline\tableline
Source Name & Observation$^a$ & $\alpha_{\rm L}$$^b$   & $\alpha$$^c$ & $f_{\rm br}$$^d$ & $\chi^2$ (dof)\\
\tableline
Mrk~421 & \asca 1998 & 0.88$\pm$0.43 & 2.14 (fixed)& (9.5$\pm$0.1)$\times$10$^{-6}$  & 21.1 (22)\\
Mrk~501 & \rxte 1998 & 1.37$\pm$0.16 & 2.92 (fixed)& (3.0$\pm$0.9)$\times$10$^{-5}$  & 18.6 (10)\\
PKS 2155$-$304 & \rxte 1996  & 1.46$\pm$0.10 & 2.23 (fixed) & (1.2$\pm$0.4)$\times$10$^{-5}$ & 3.9 (10)\\
\tableline
\end{tabular}
\tablenotetext{a}{GIS data (0.7--10 keV) and SIS data (0.5--10 keV) were 
used for \asca observations and PCA data (2.5--20 keV) were used for 
\rxte observations.}
\tablenotetext{b}{The best fit broken power-law index of NPSD below
the break frequency $f$ $<$ $f_{\rm br}$.}
\tablenotetext{c}{The best fit power-low index of NPSDs in the region 
of $10^{-5}$ Hz $<$ $f$ $<$ 10$^{-3}$ Hz.}
\tablenotetext{d}{The best fit break frequency $f_{\rm br}$.}
\end{center}
\end{table}

\clearpage

\begin{figure}
\epsscale{0.8}
\plotone{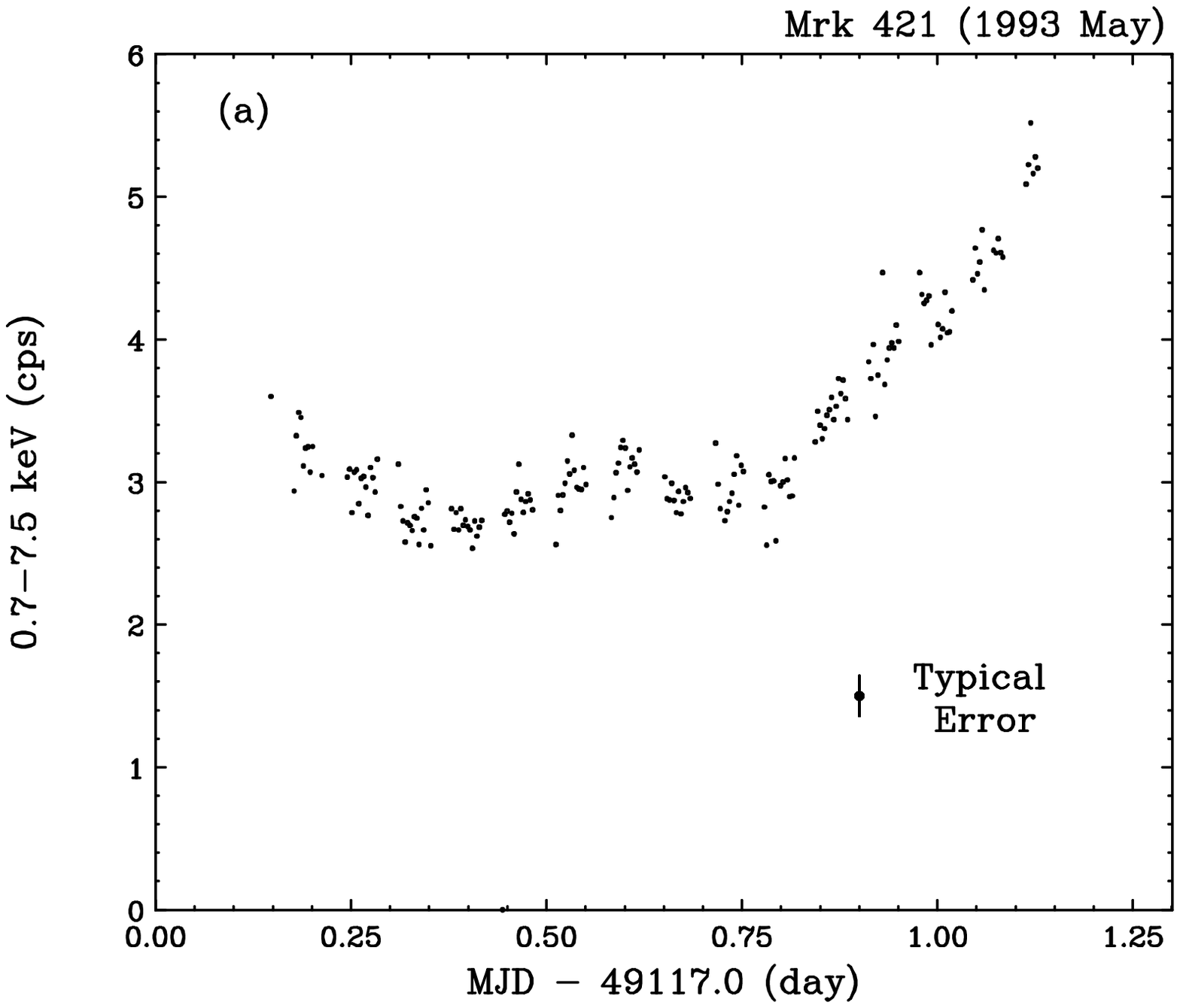}
\plotone{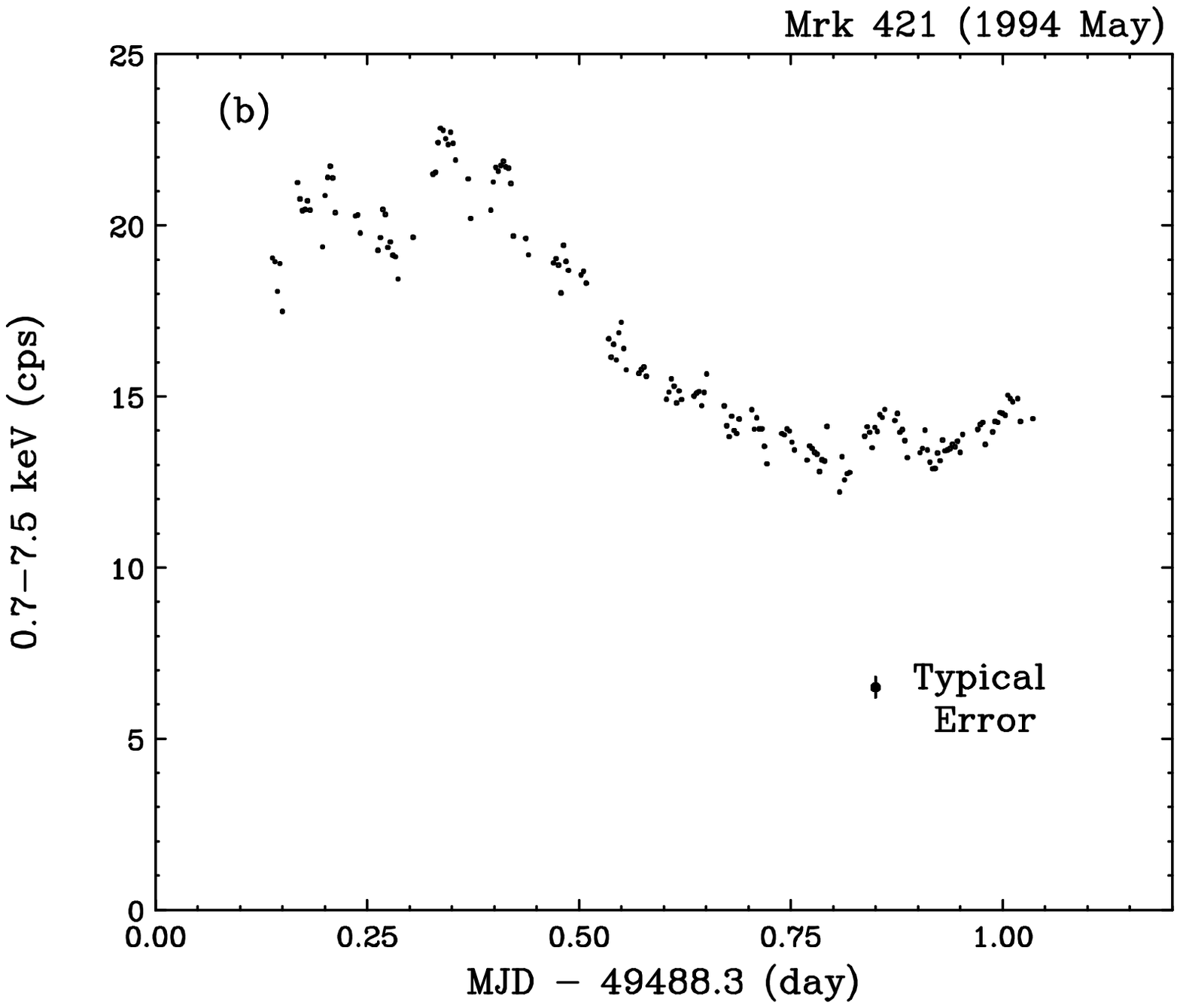}
\end{figure}
\clearpage

\begin{figure}
\epsscale{0.8}
\plotone{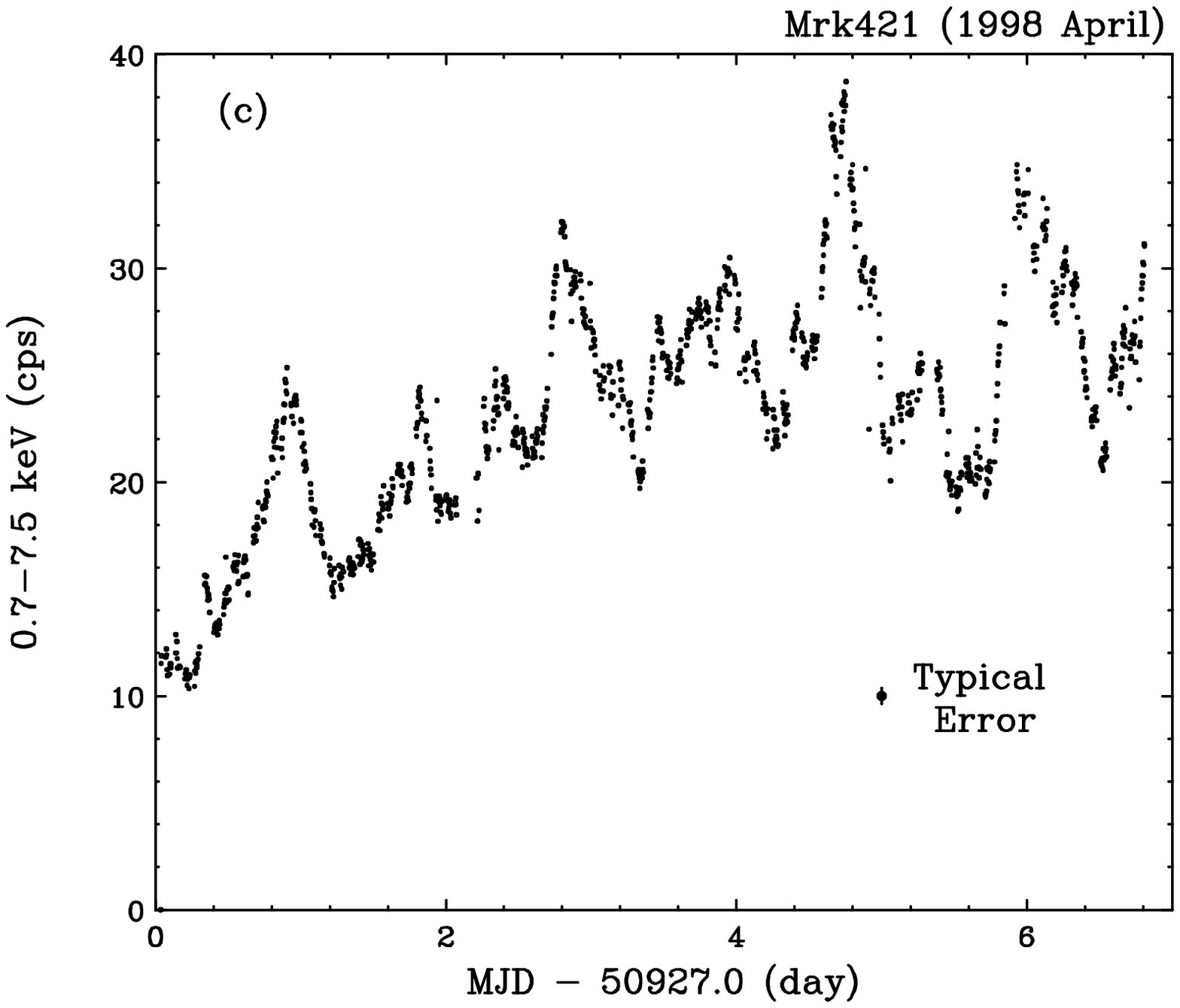}
\plotone{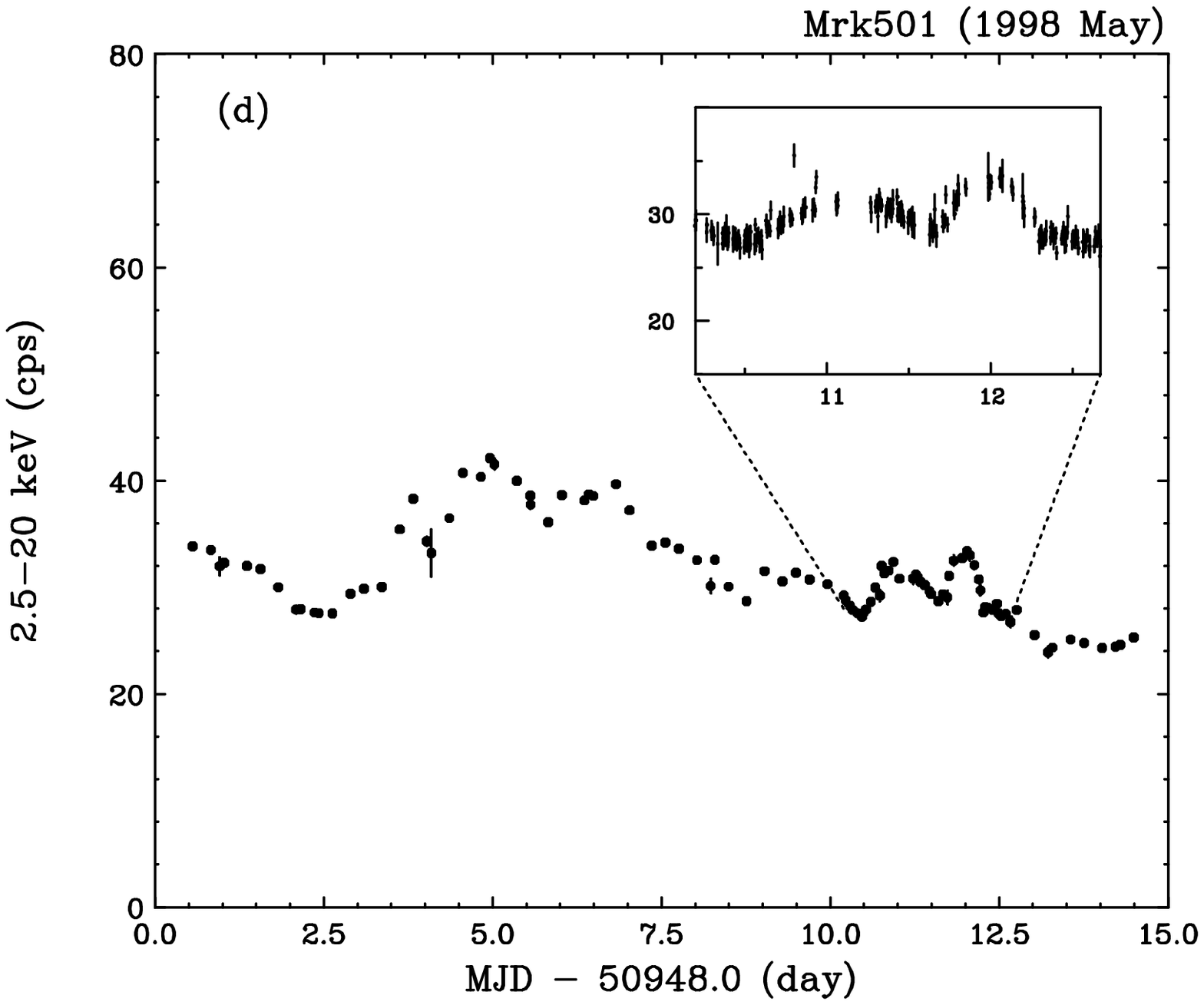}
\end{figure}
\clearpage

\begin{figure}
\epsscale{0.8}
\plotone{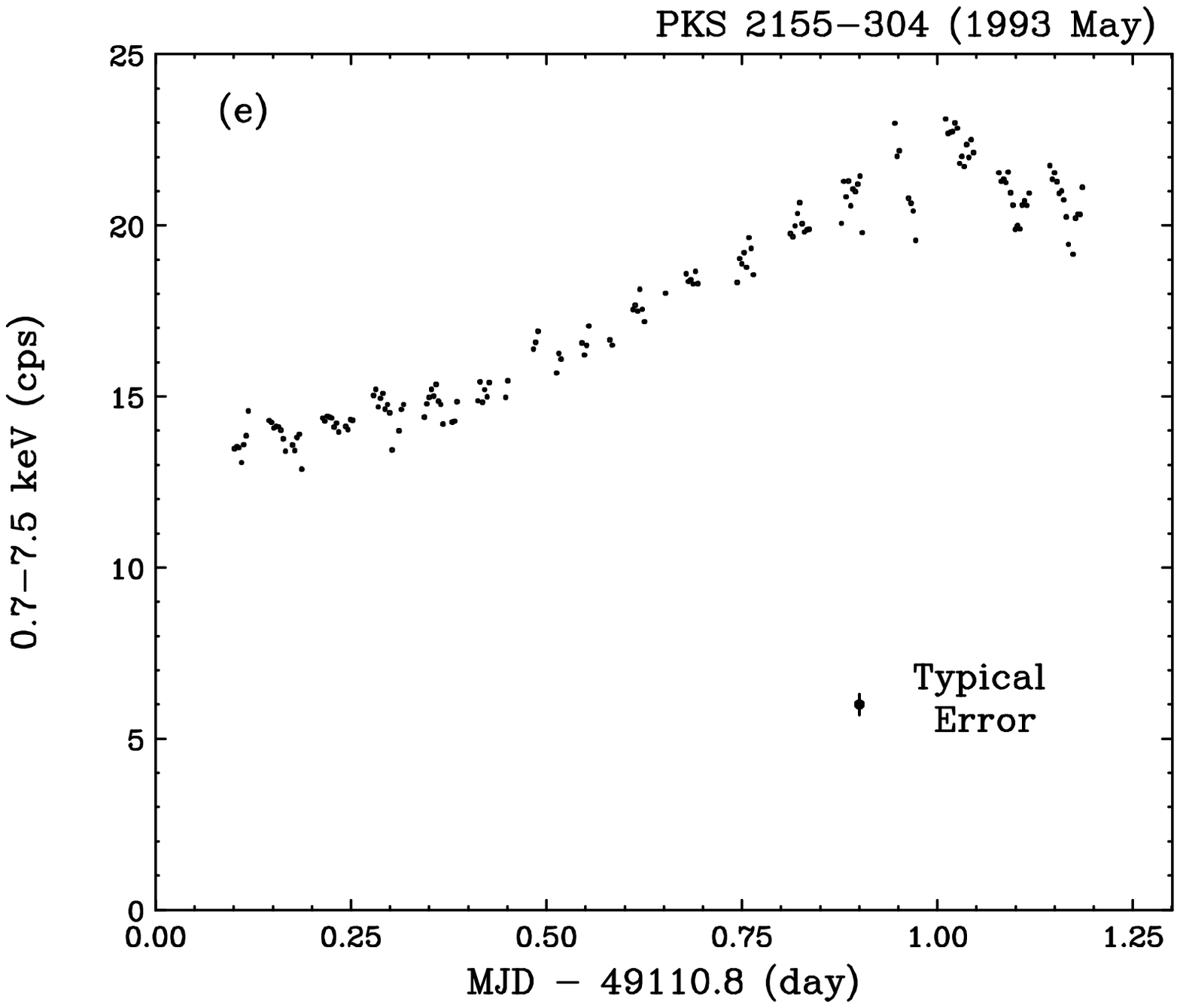}
\plotone{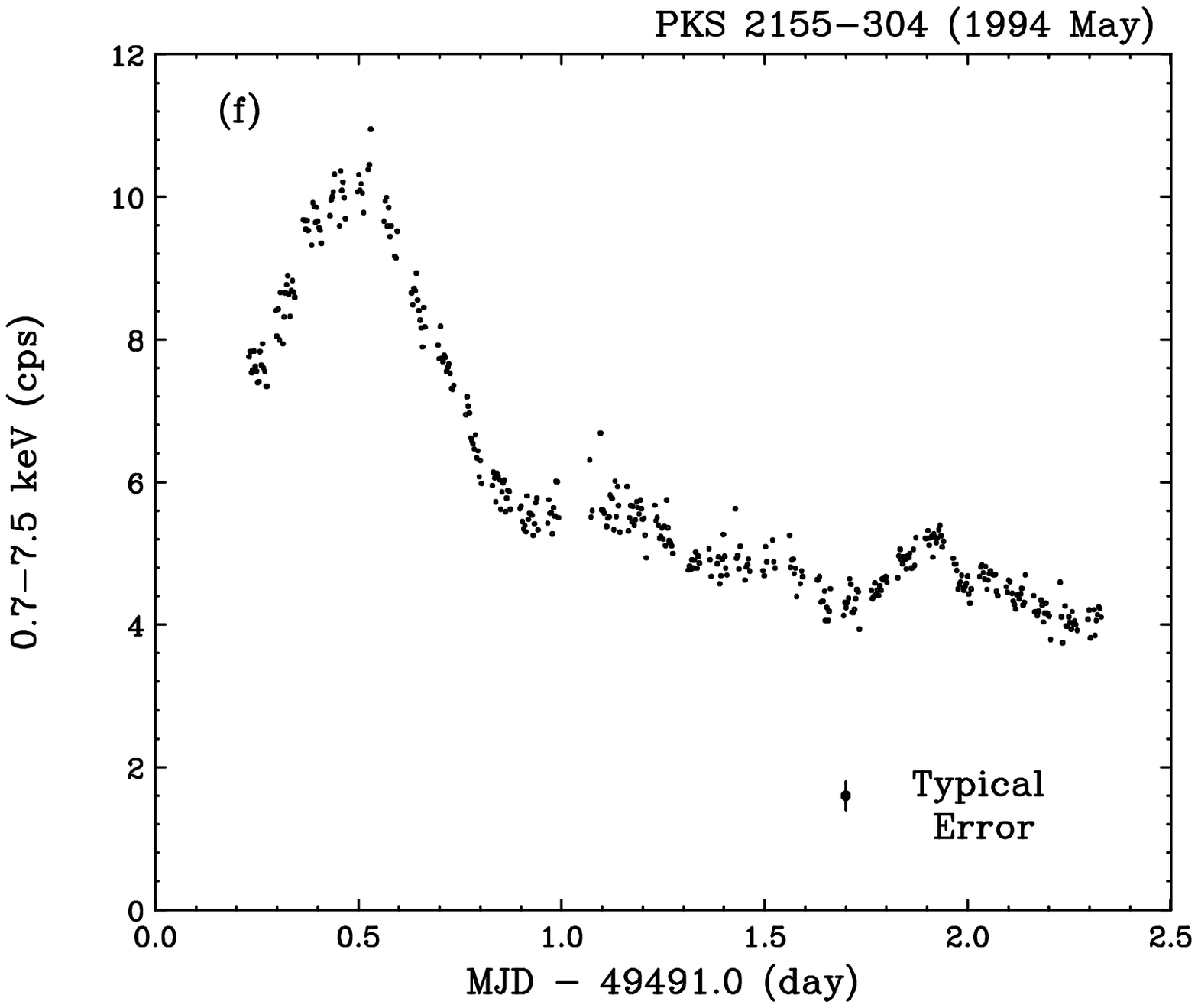}
\end{figure}
\clearpage

\begin{figure}
\epsscale{0.8}
\plotone{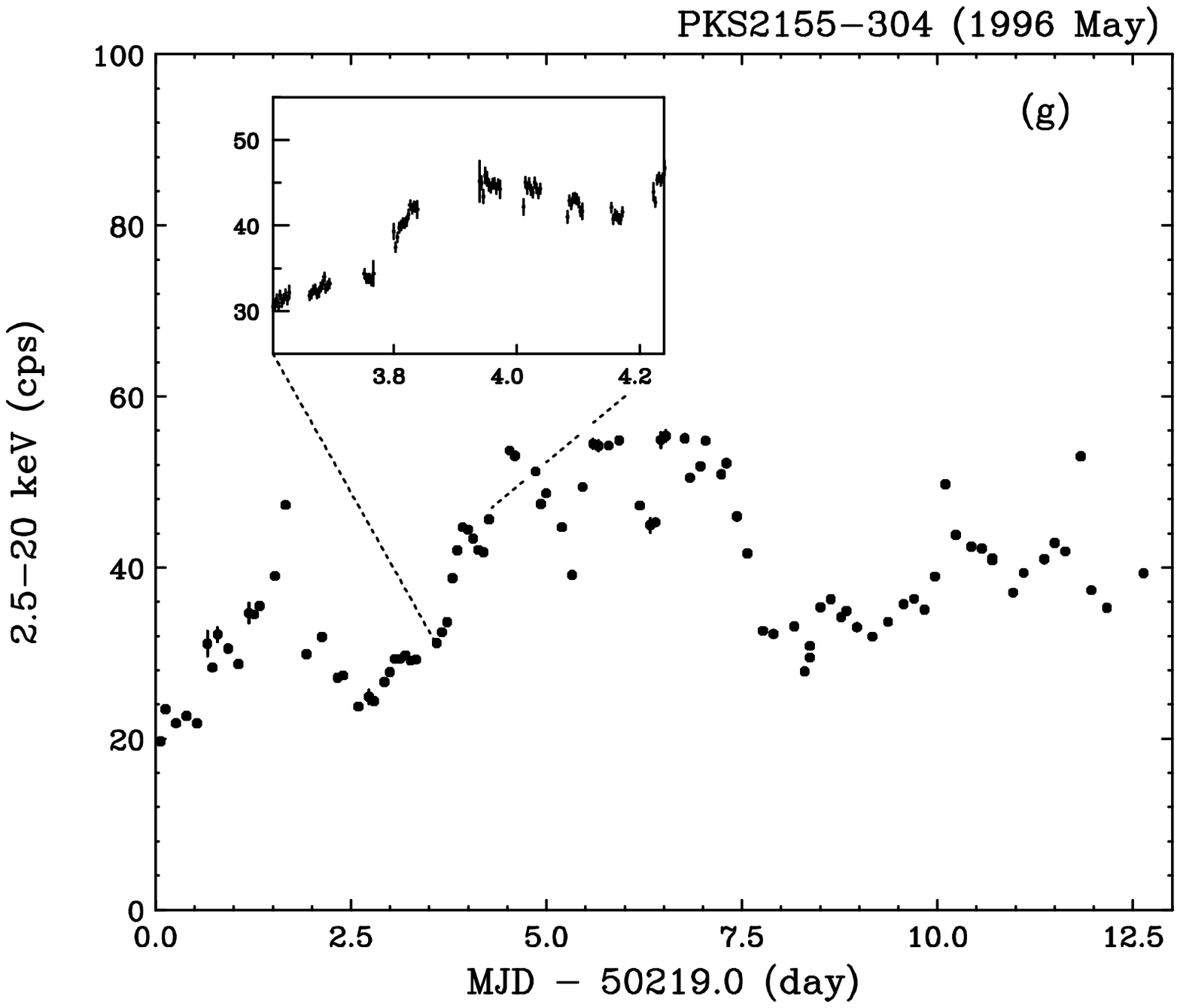}
\end{figure}
\clearpage

\begin{figure}
\epsscale{0.8}
\plotone{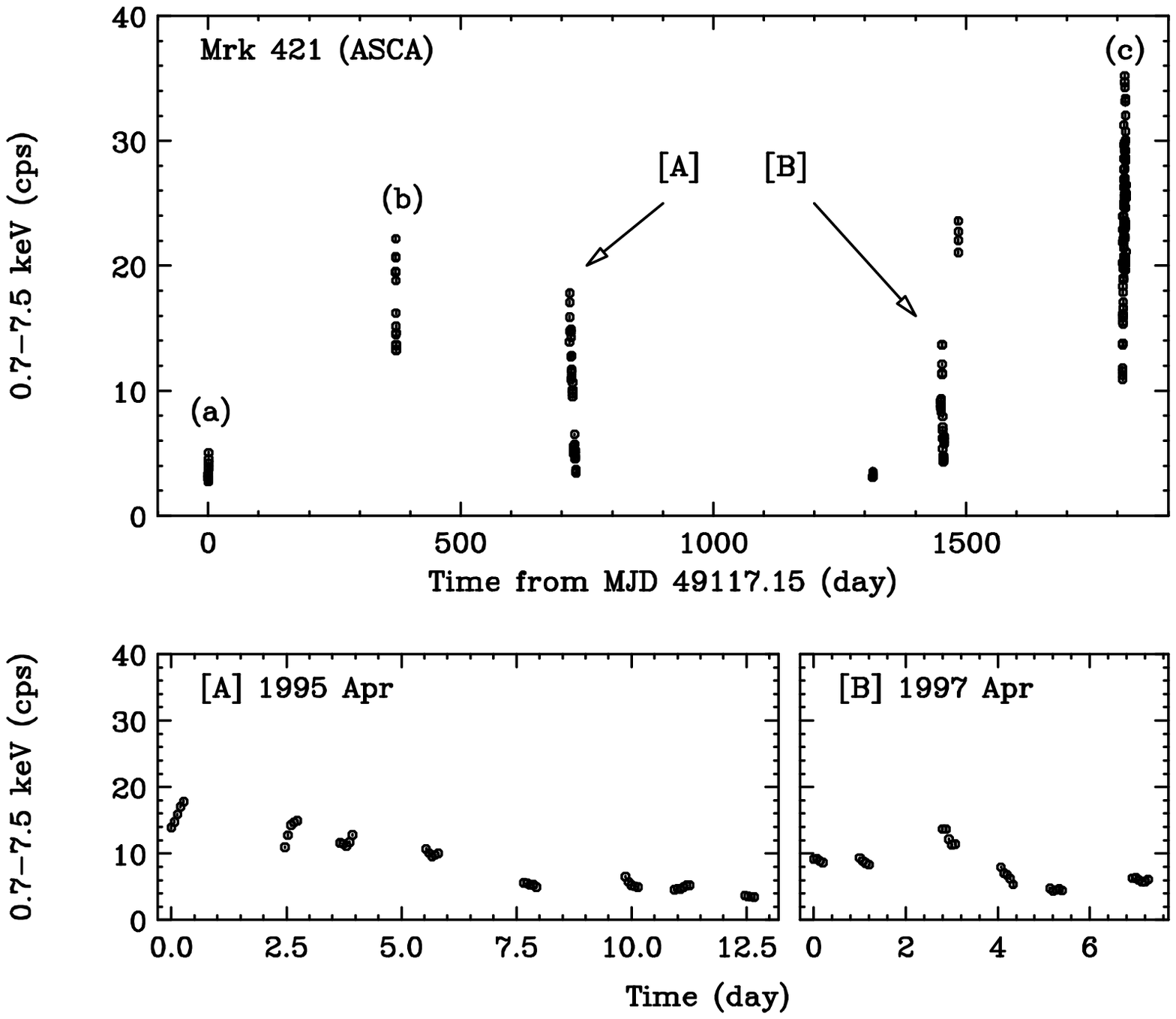}
\end{figure}

\clearpage
\begin{figure}
\epsscale{0.8}
\plotone{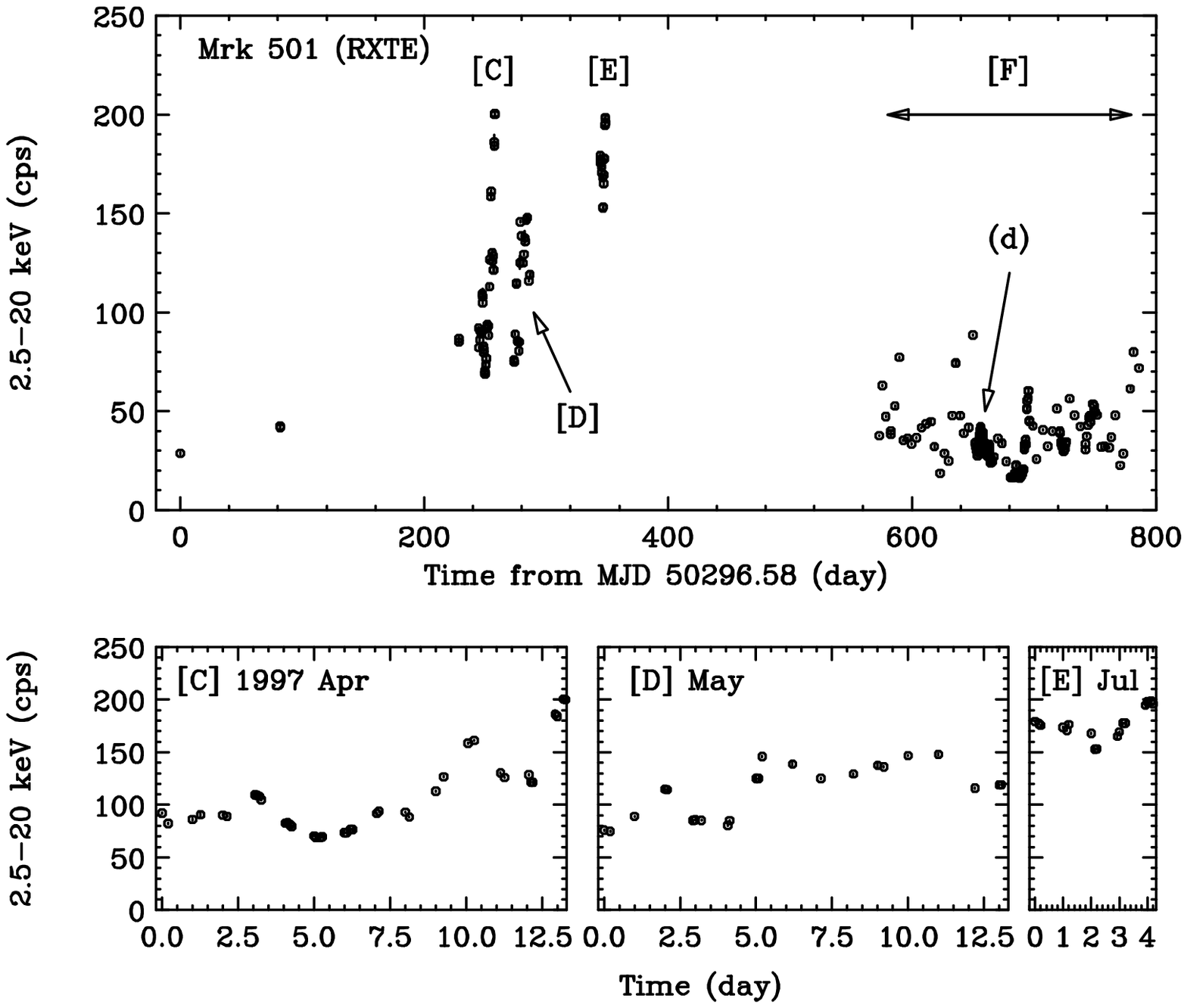}
\end{figure}

\clearpage

\begin{figure}
\epsscale{0.8}
\plotone{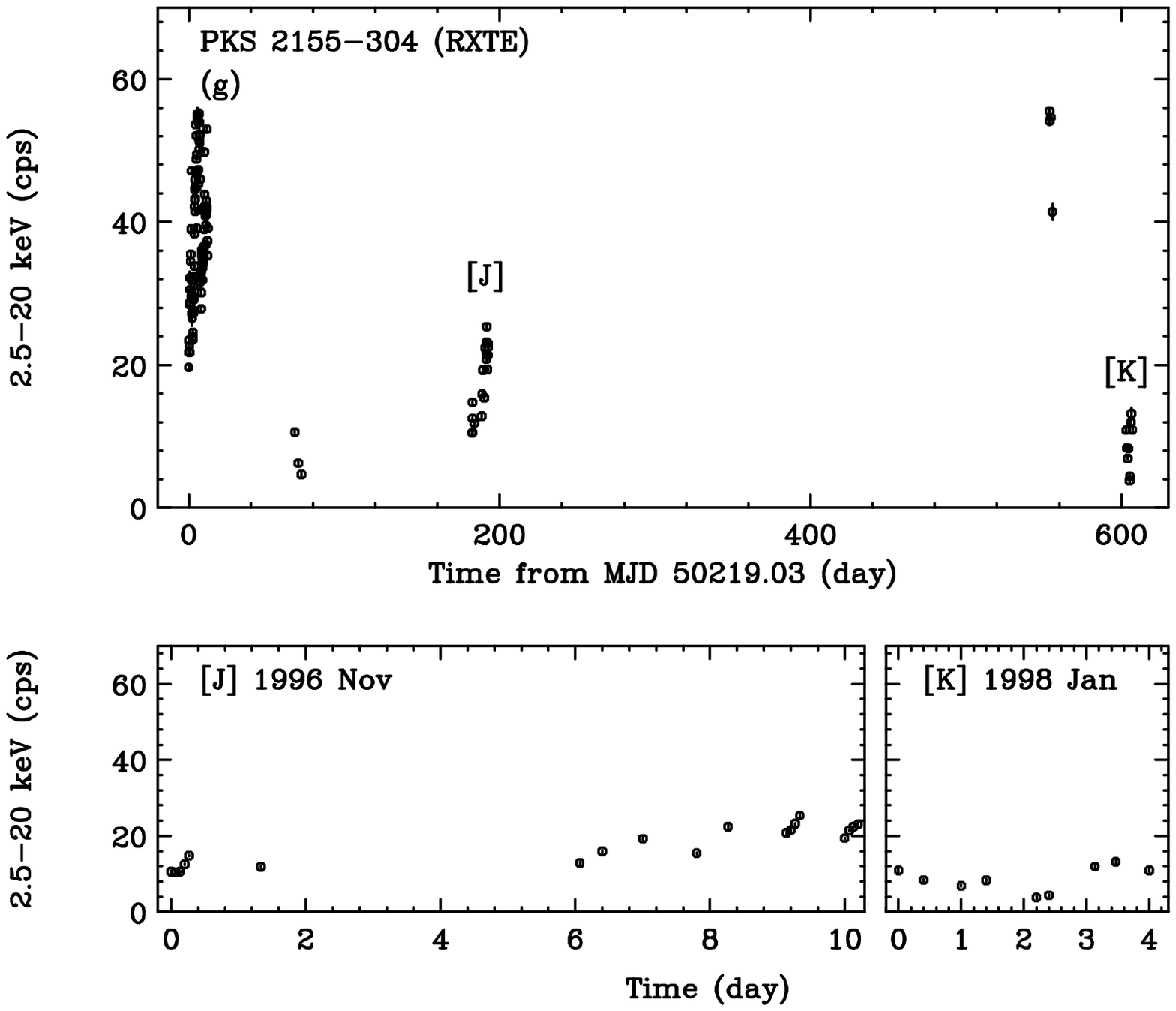}
\end{figure}

\clearpage

\begin{figure}
\epsscale{0.8}
\plotone{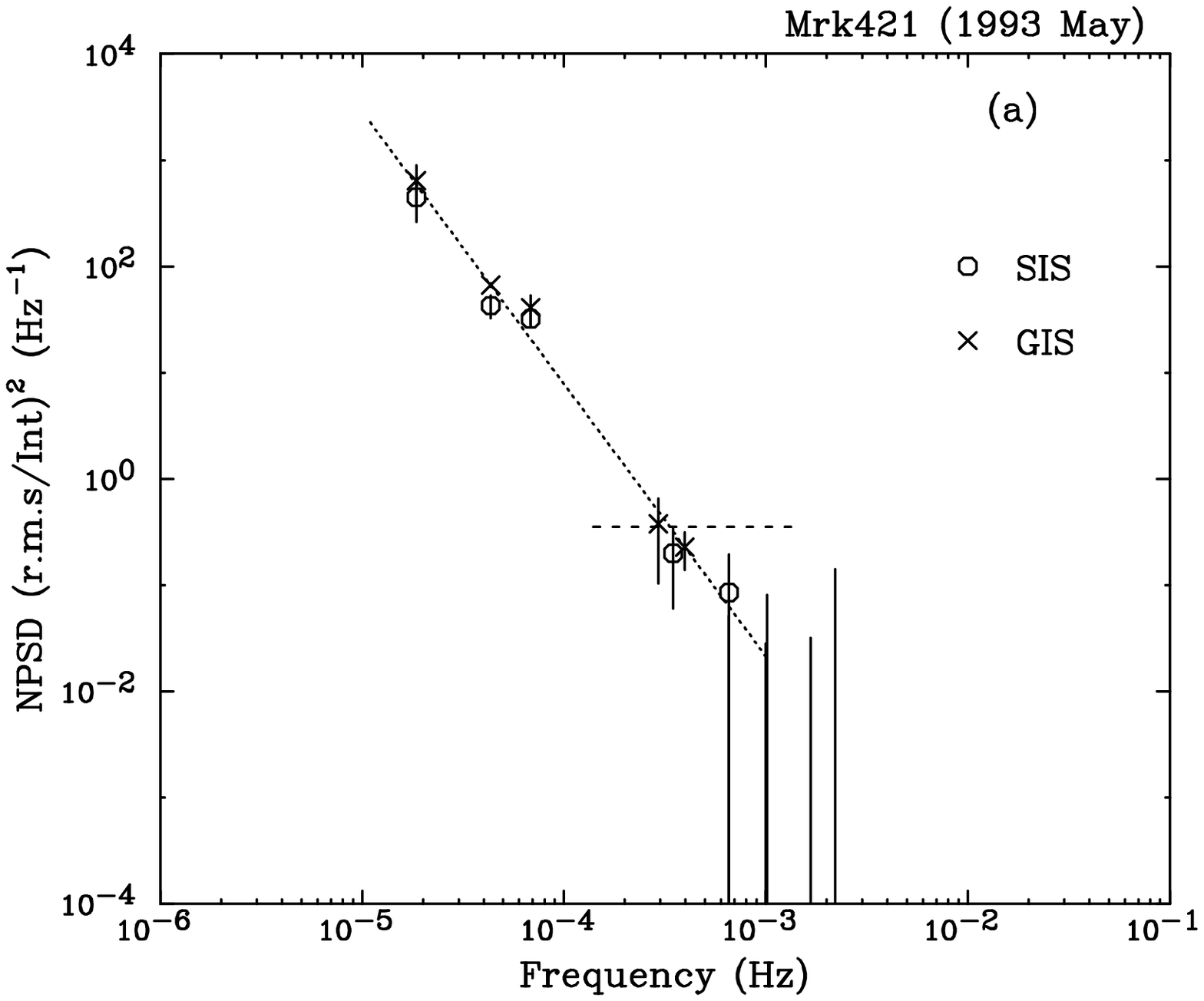}
\plotone{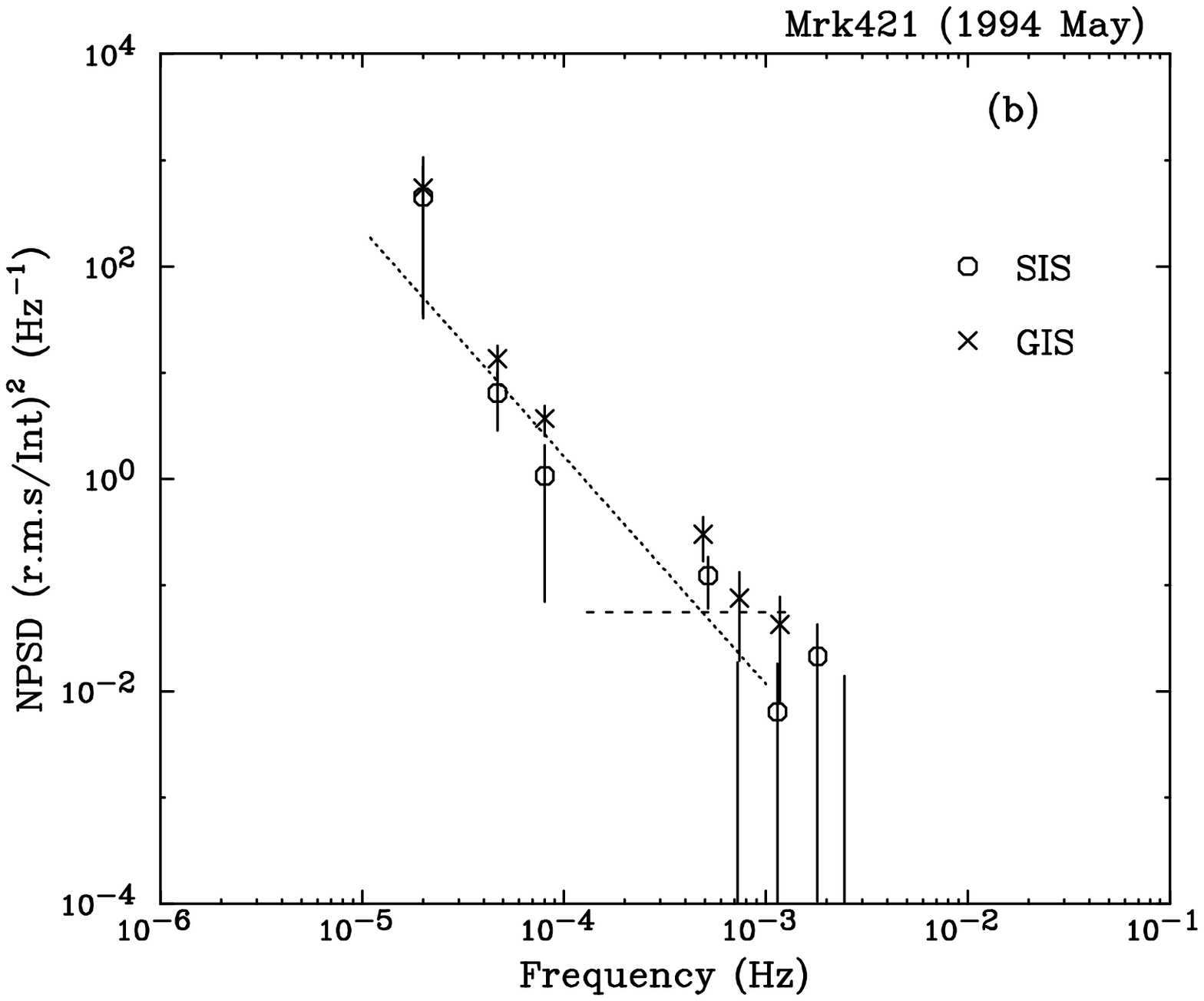}
\end{figure}

\clearpage

\begin{figure}
\epsscale{0.8}
\plotone{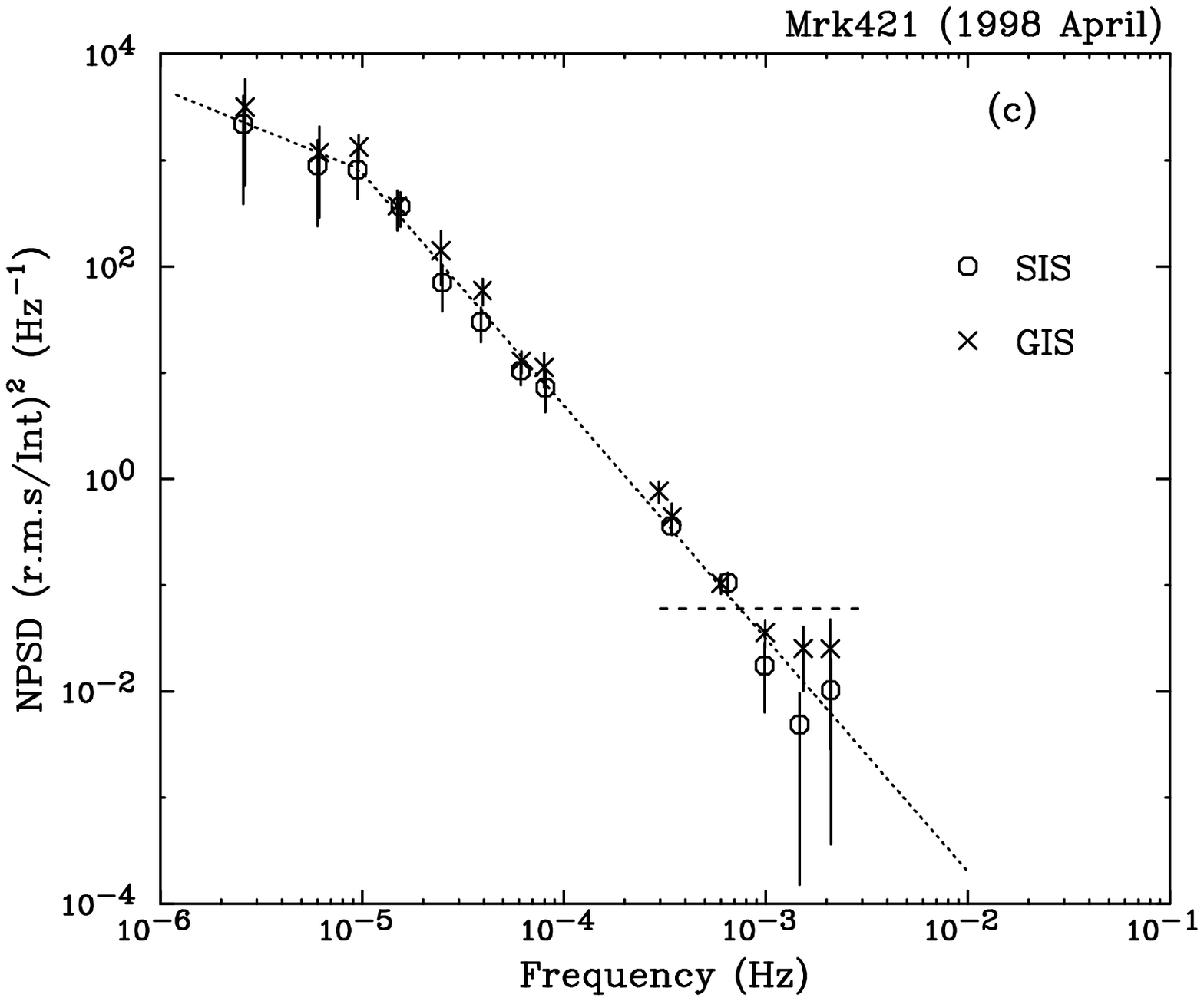}
\plotone{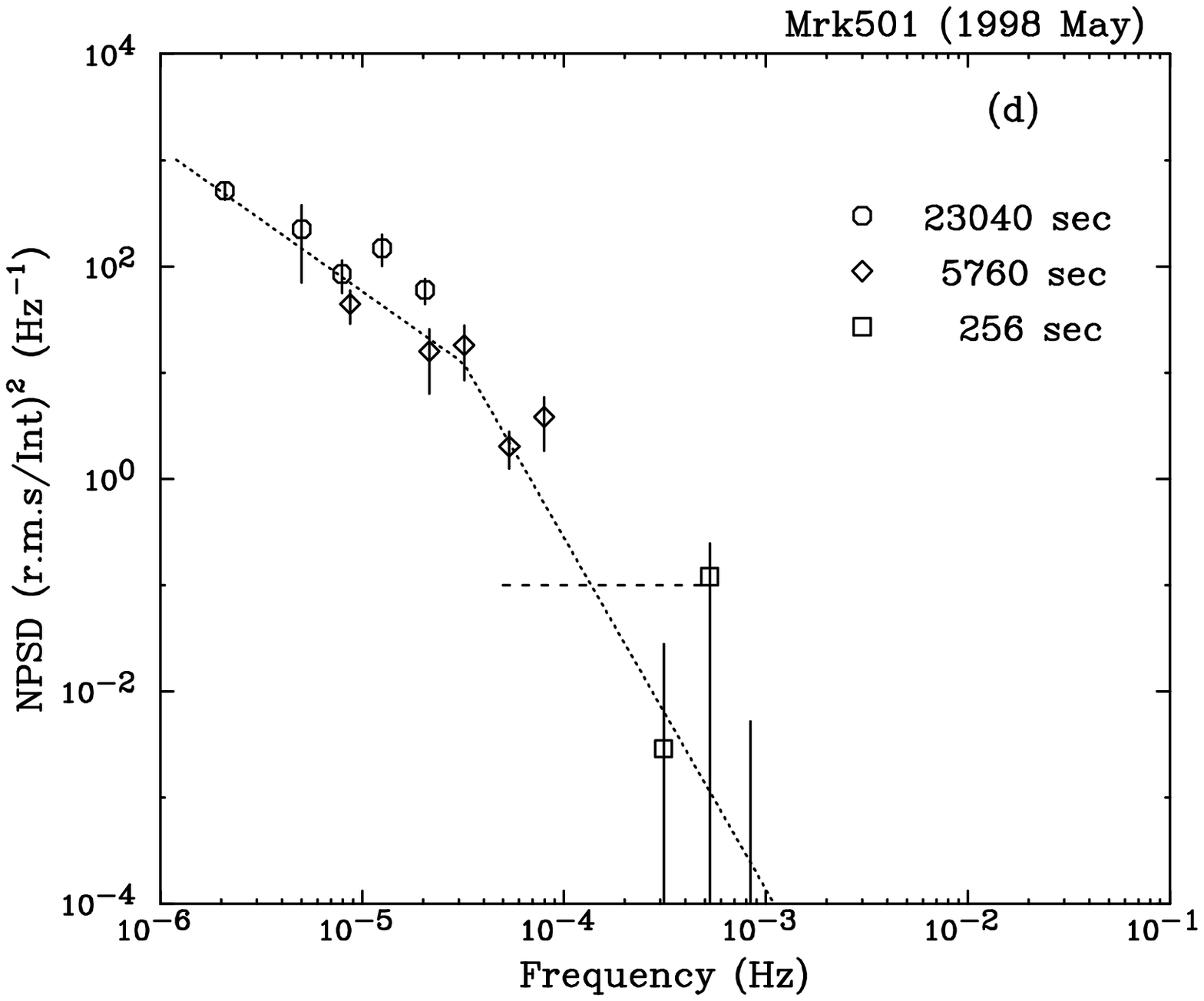}
\end{figure}
\clearpage

\begin{figure}
\epsscale{0.8}
\plotone{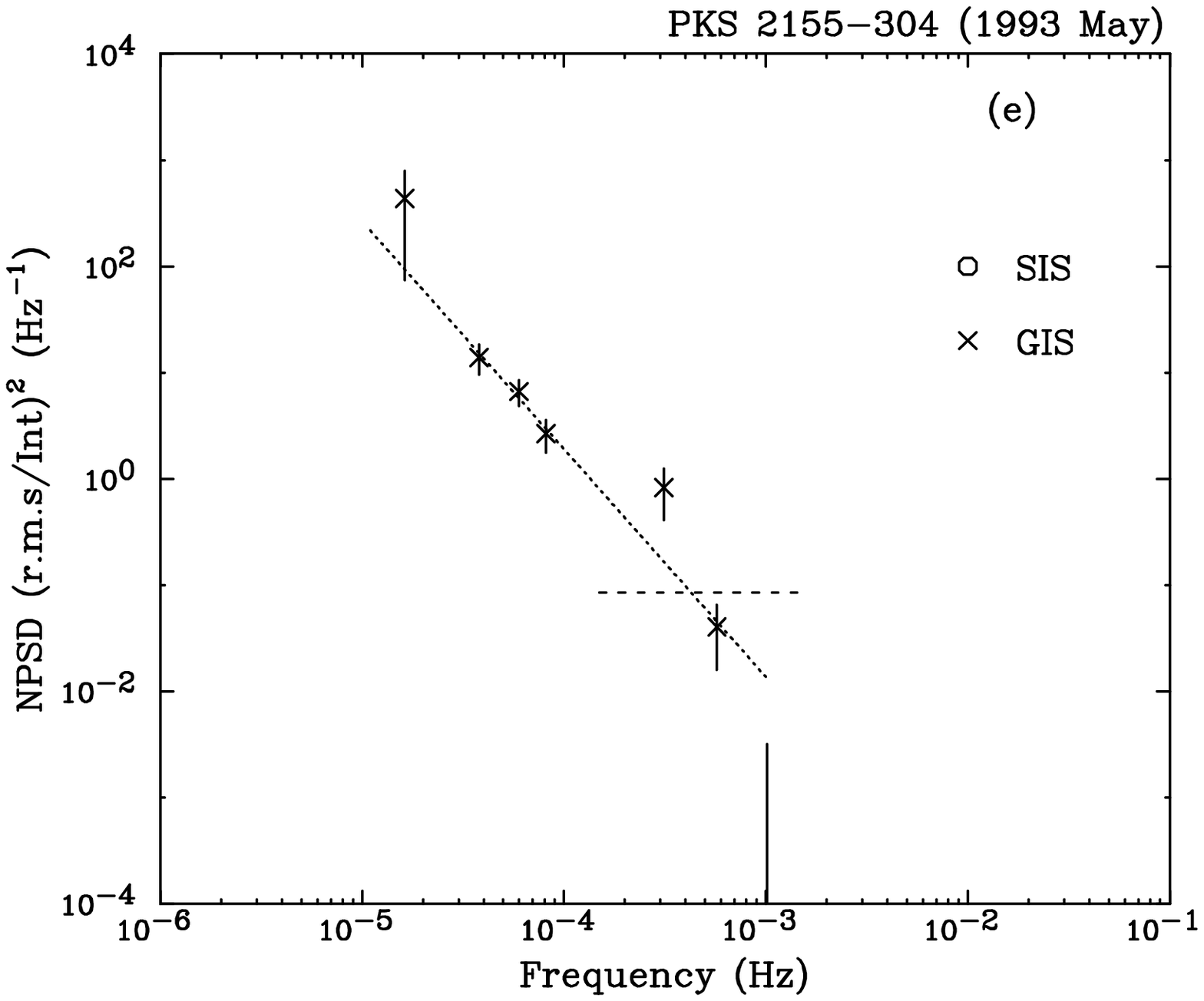}
\plotone{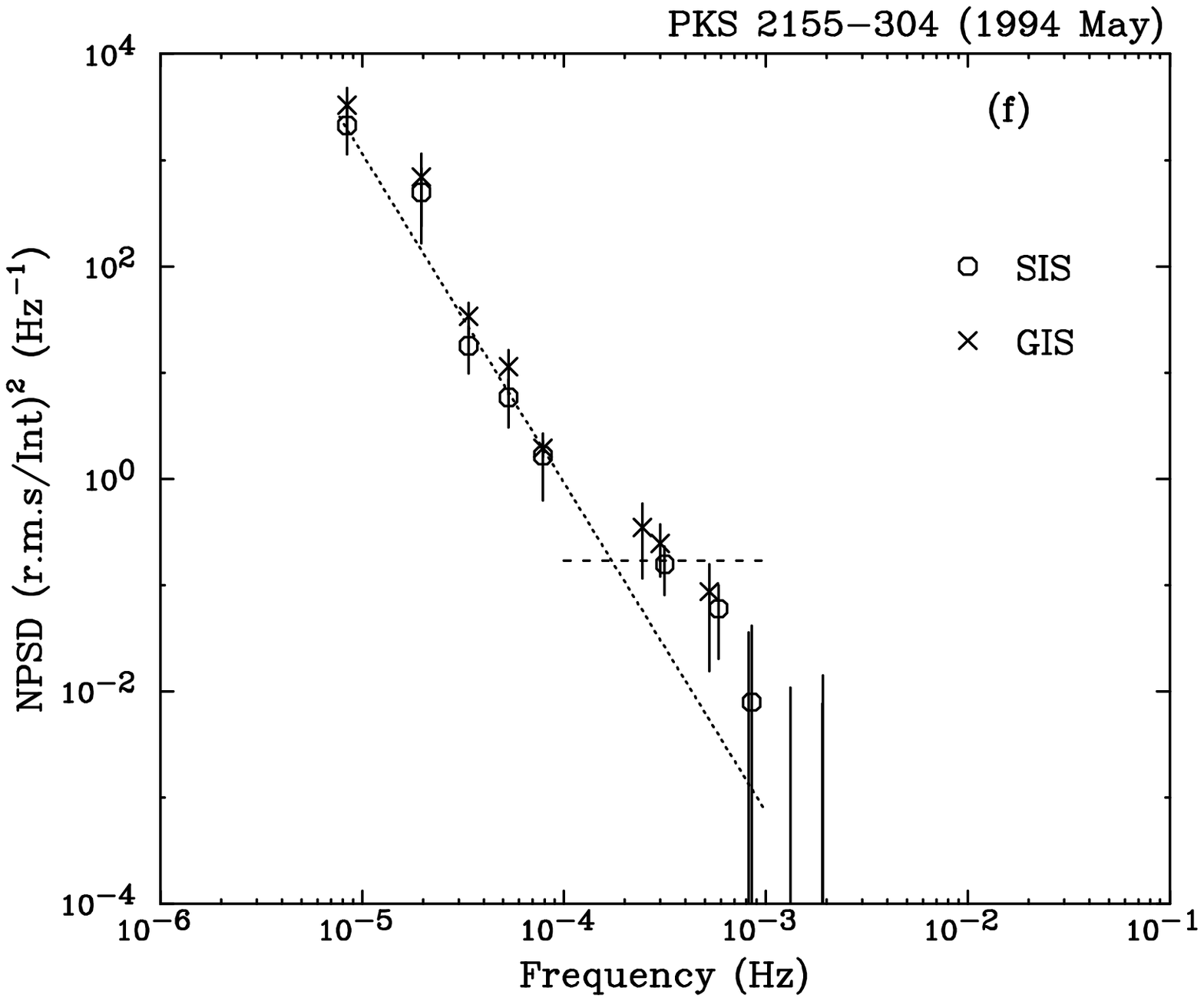}
\end{figure}
\clearpage

\begin{figure}
\epsscale{0.8}
\plotone{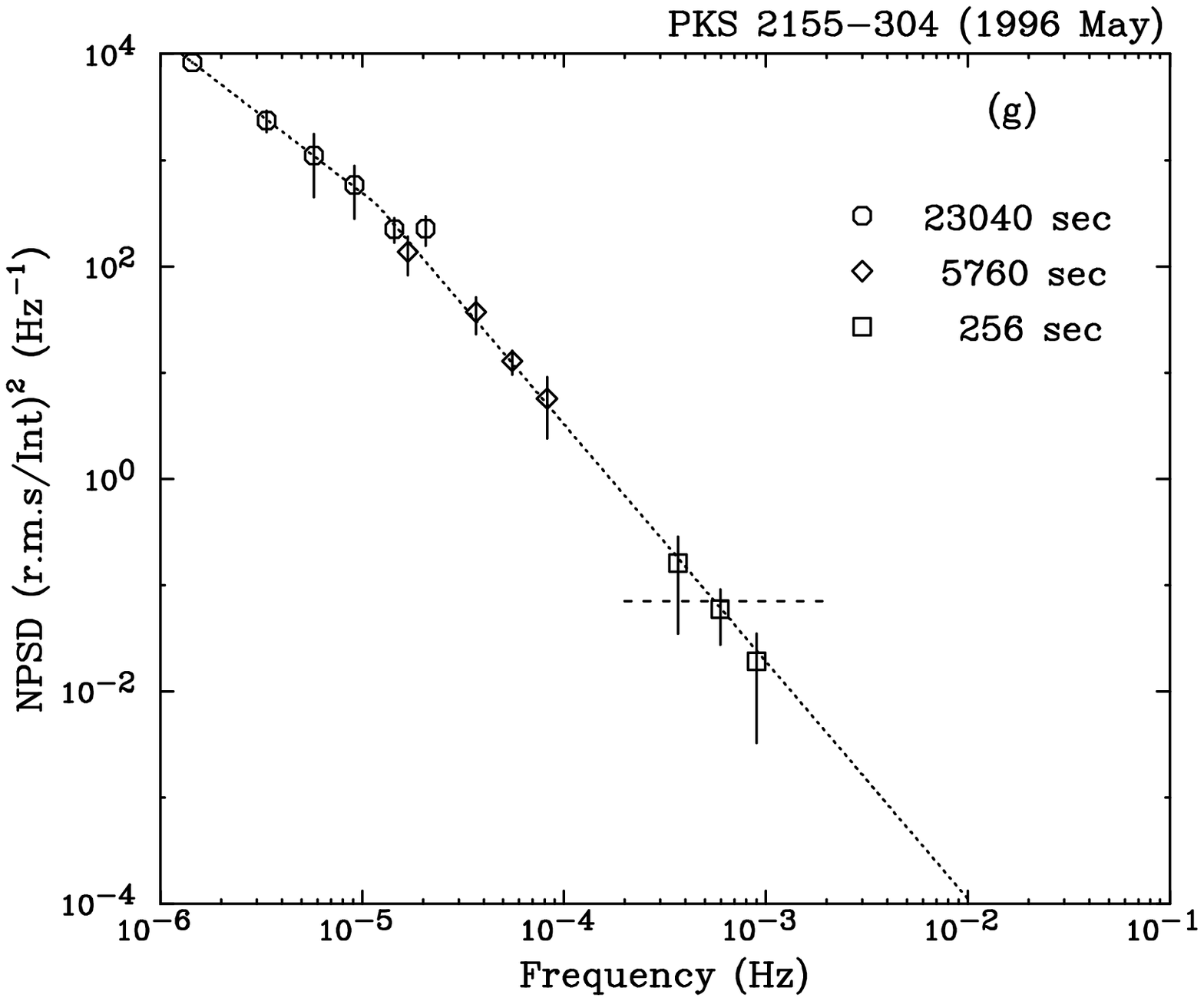}
\end{figure}
\clearpage

\begin{figure}
\epsscale{0.8}
\plotone{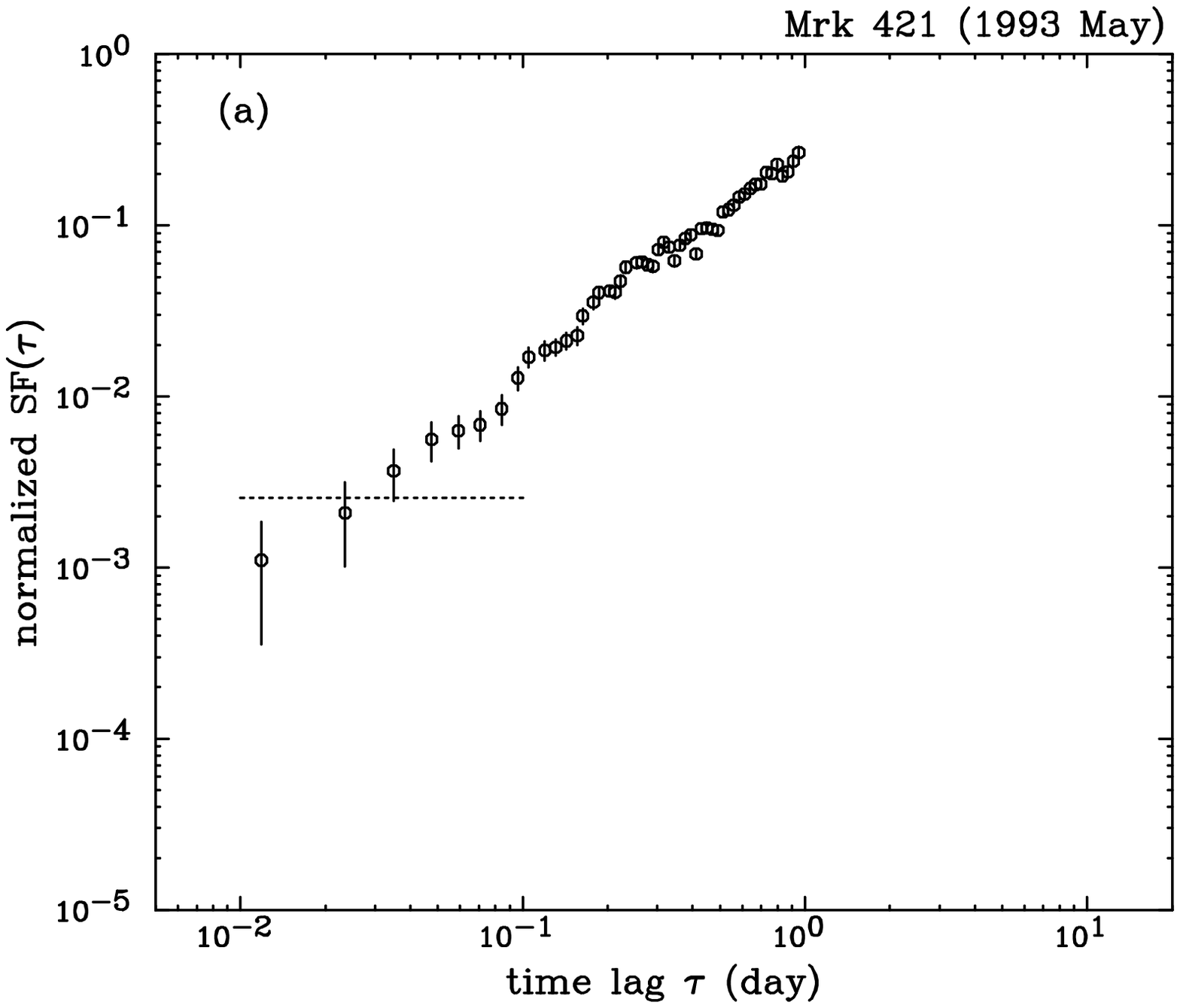}
\plotone{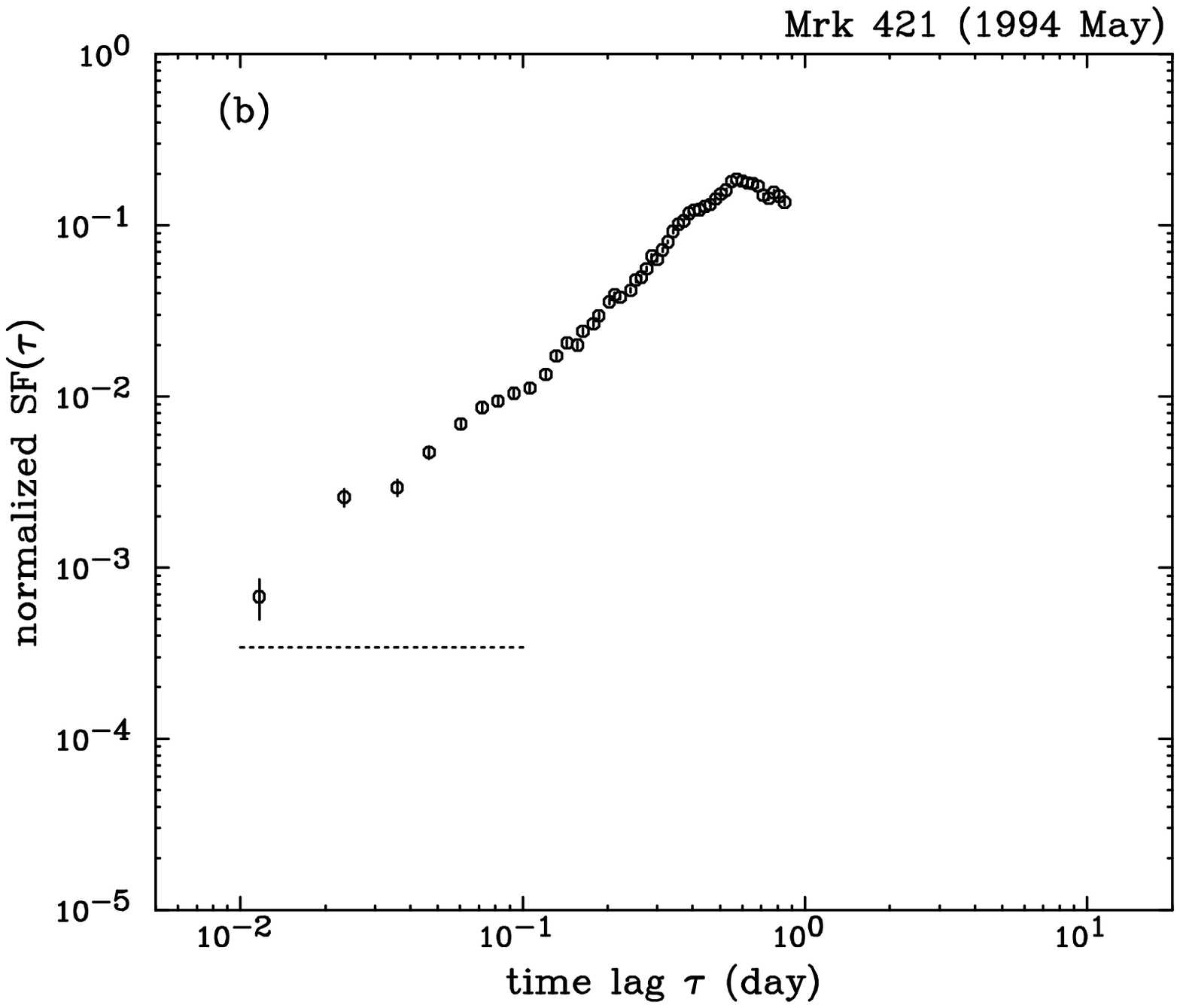}
\end{figure}
\clearpage

\begin{figure}
\epsscale{0.8}
\plotone{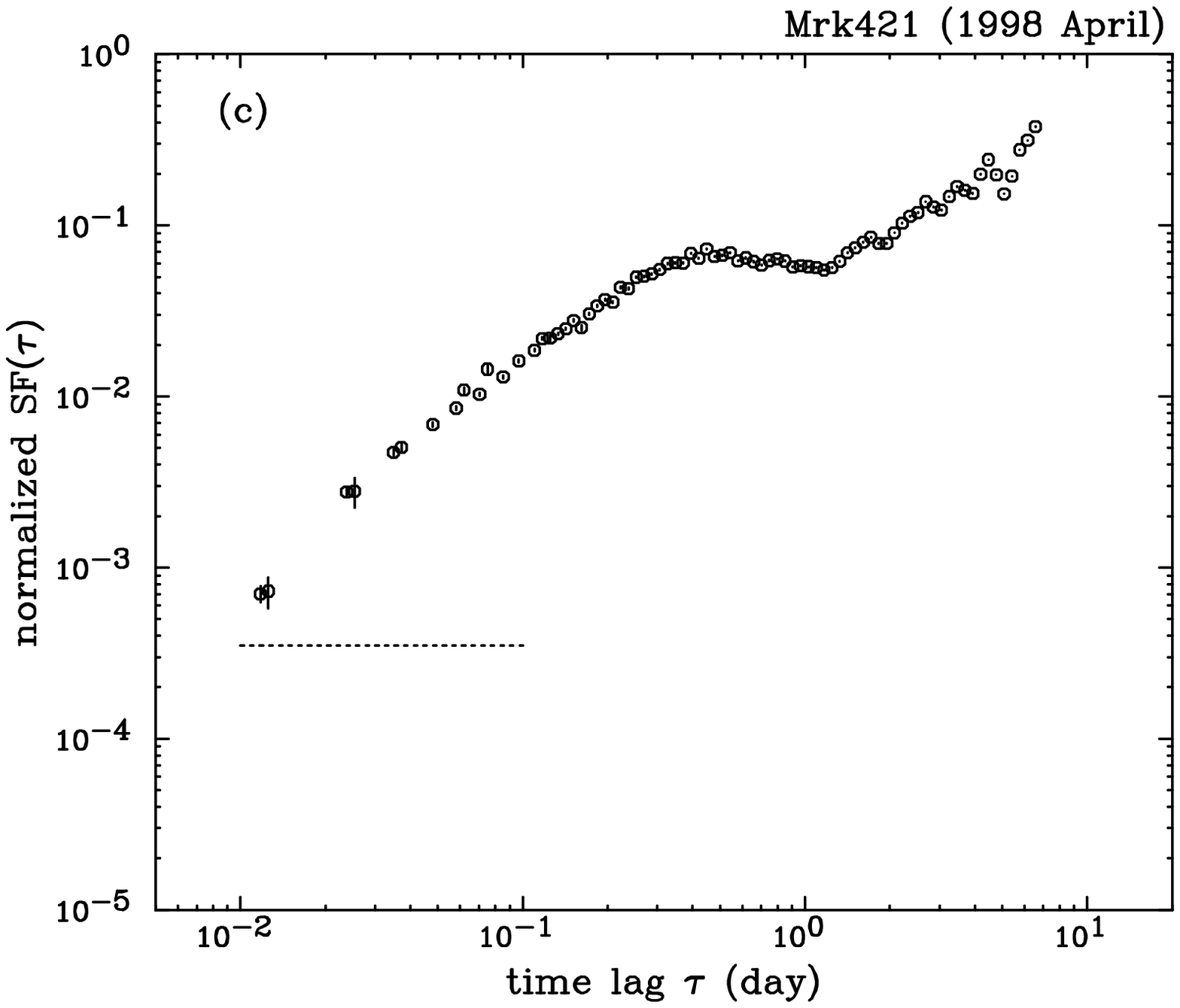}
\plotone{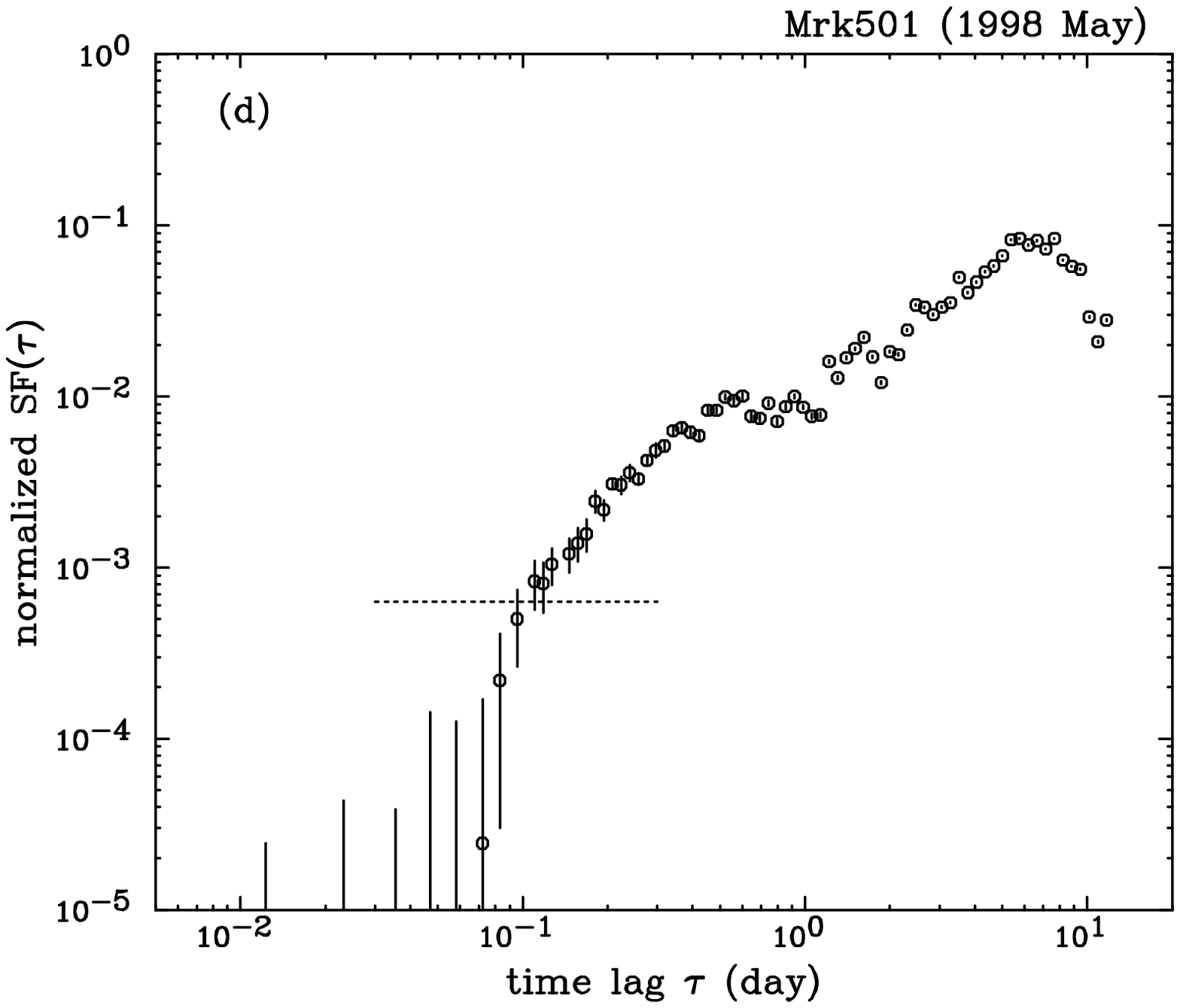}
\end{figure}
\clearpage

\begin{figure}
\epsscale{0.8}
\plotone{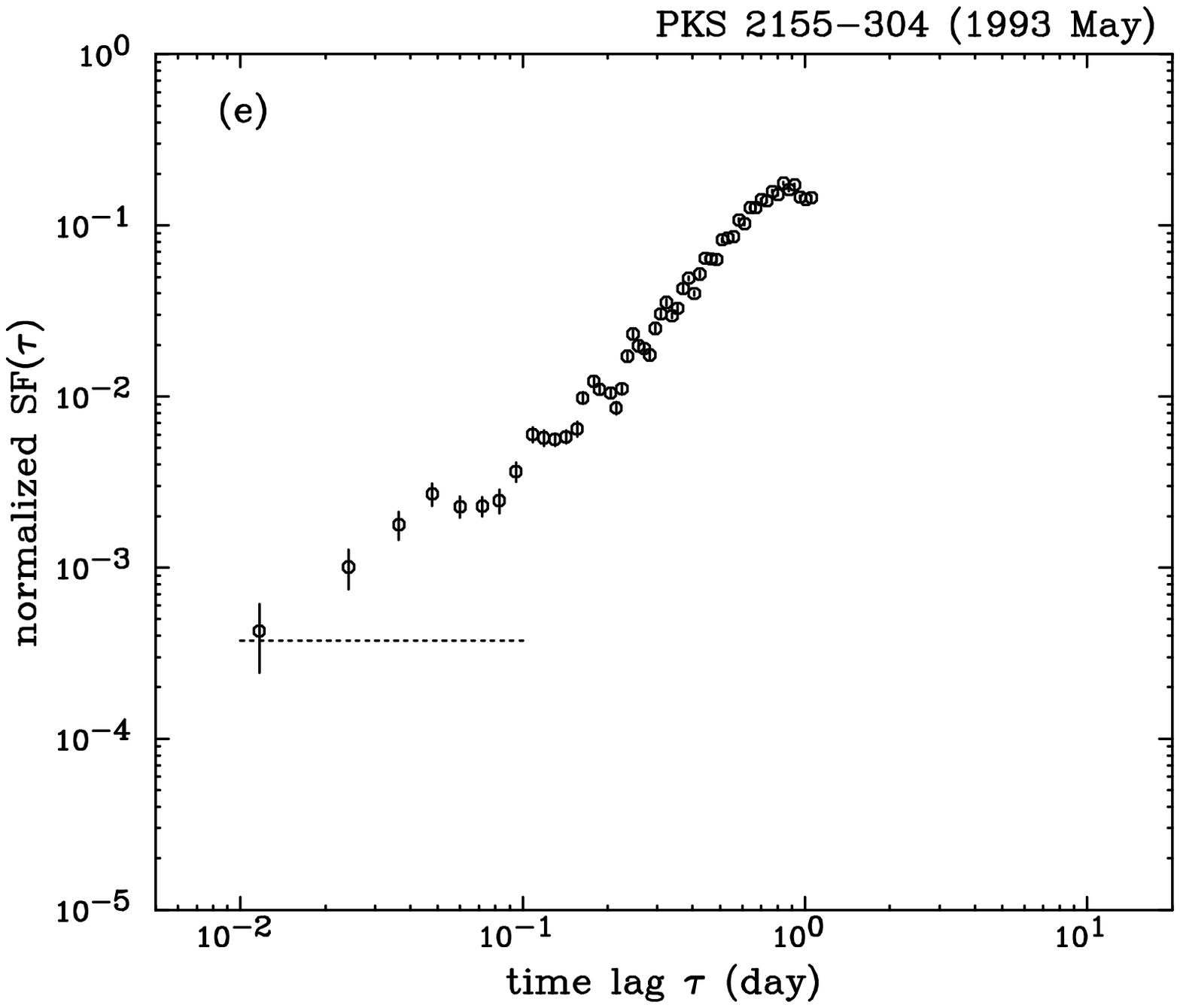}
\plotone{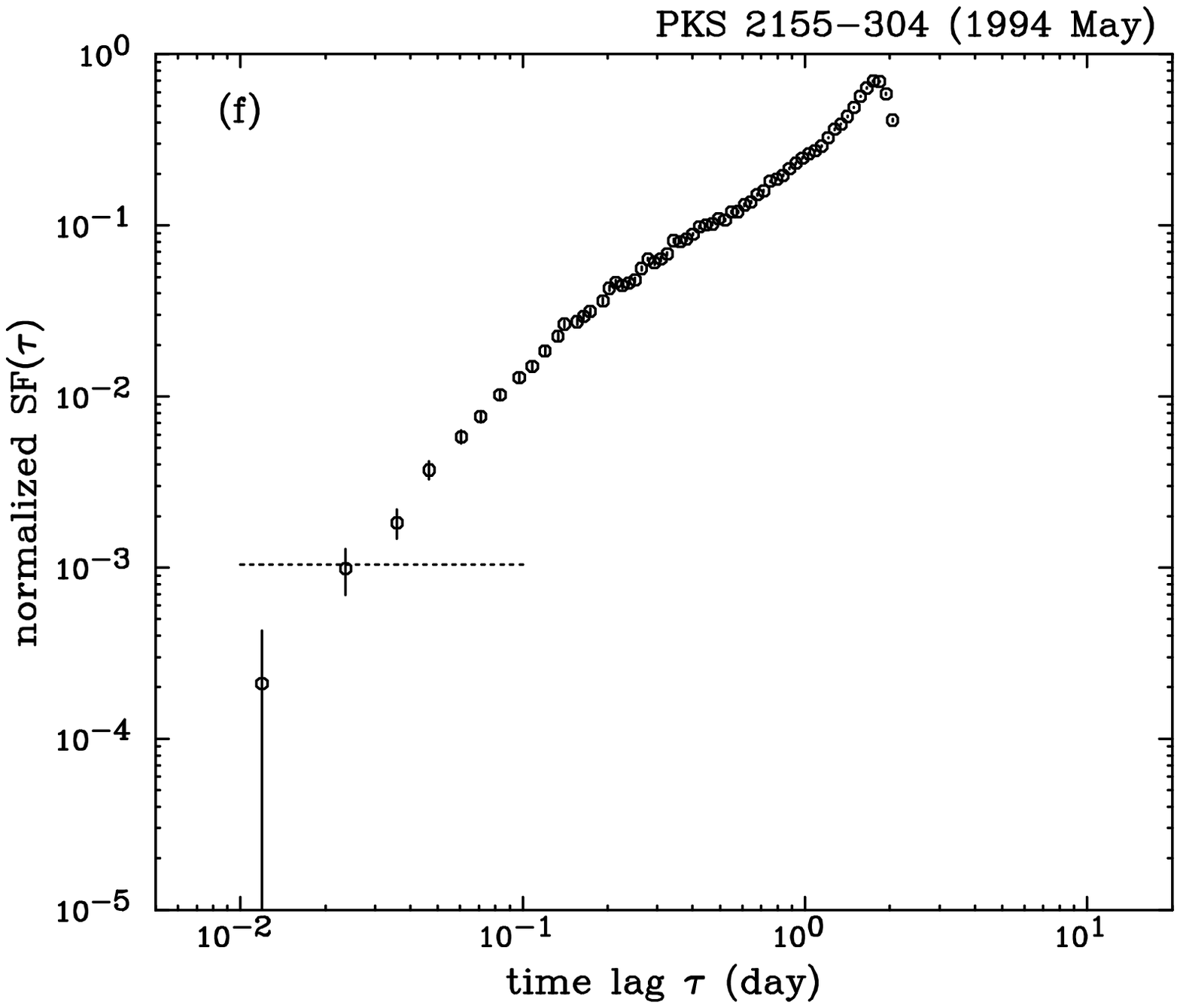}
\end{figure}
\clearpage

\begin{figure}
\epsscale{0.8}
\plotone{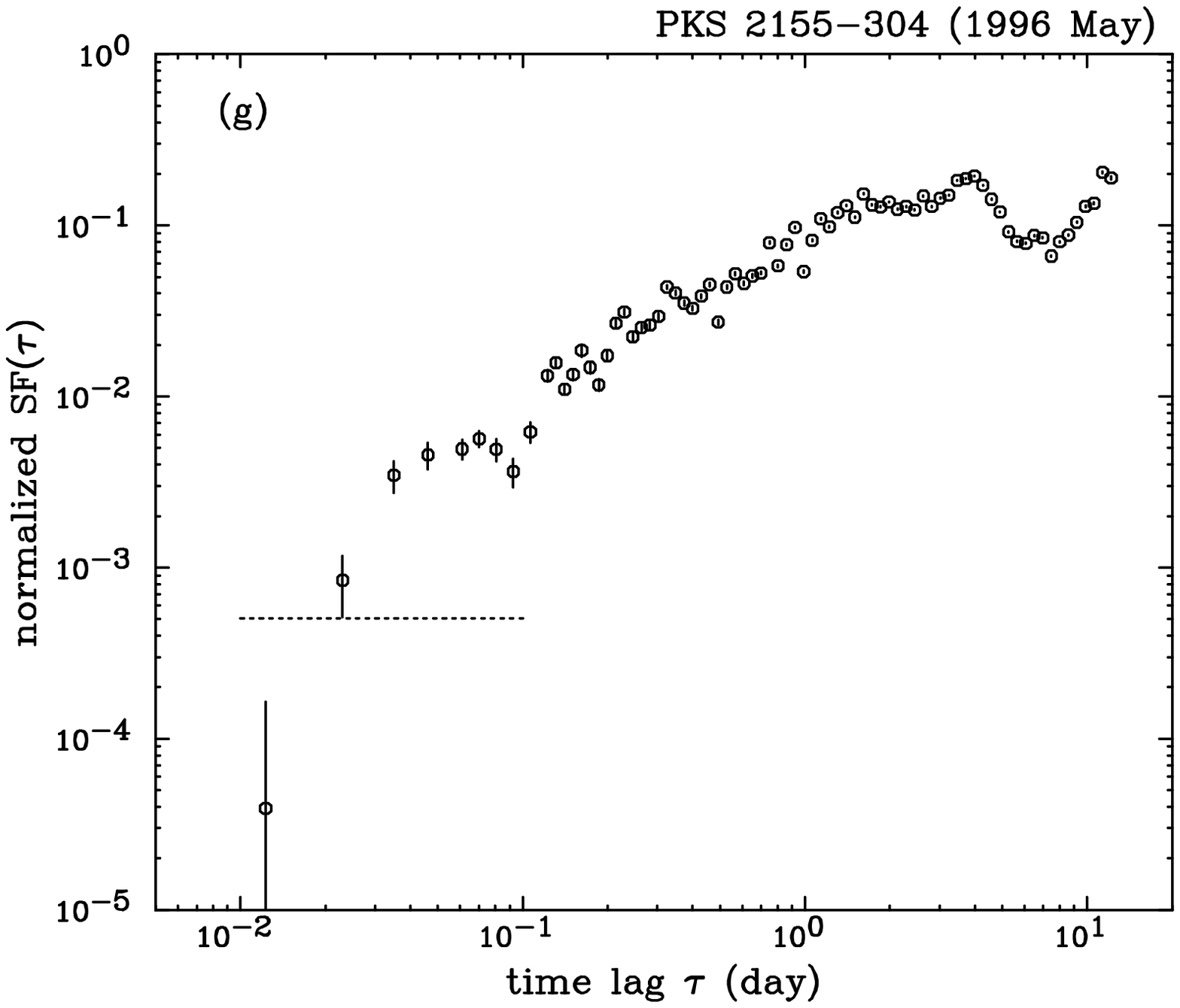}
\end{figure}
\clearpage

\begin{figure}
\epsscale{0.8}
\plotone{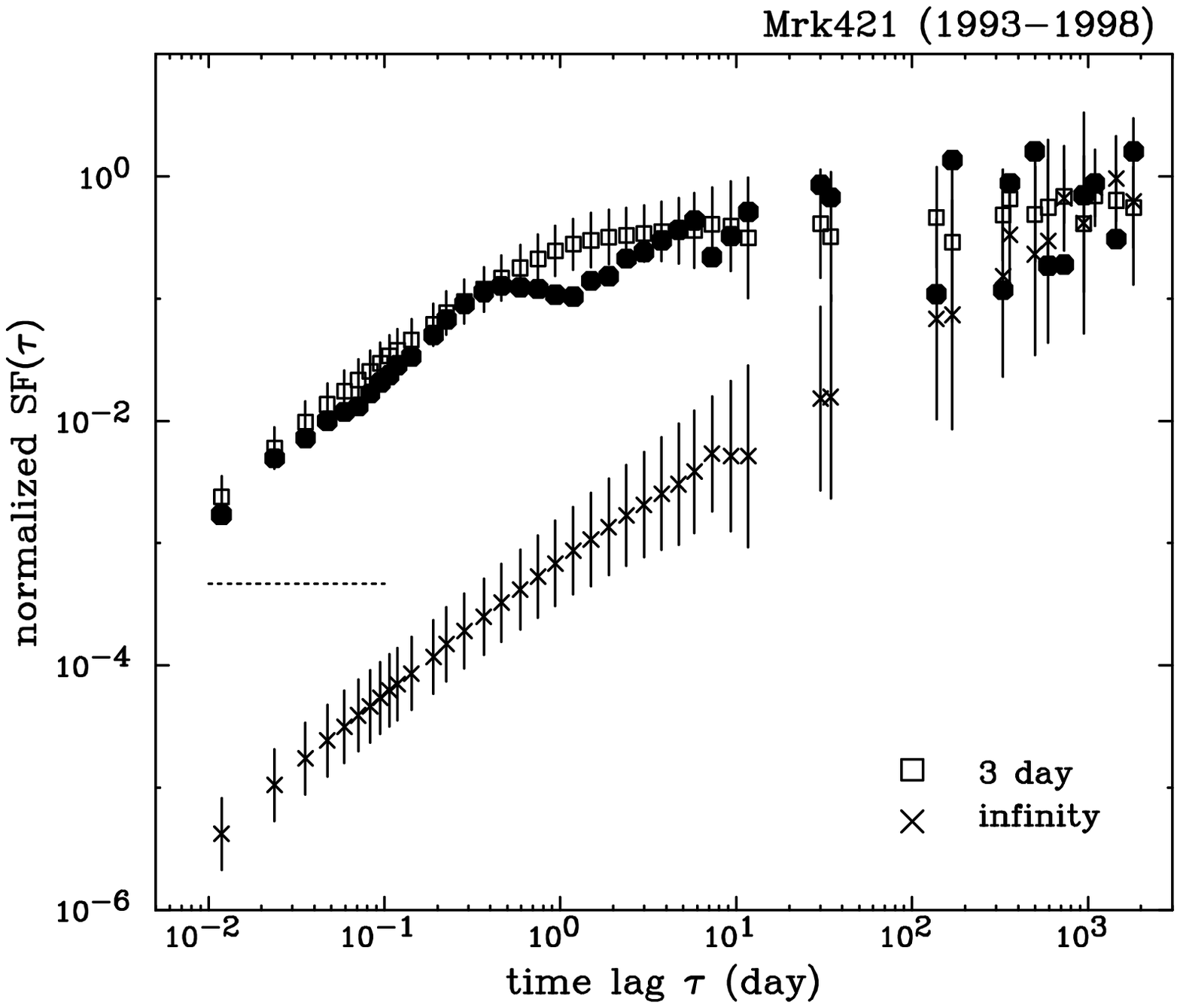}
\plotone{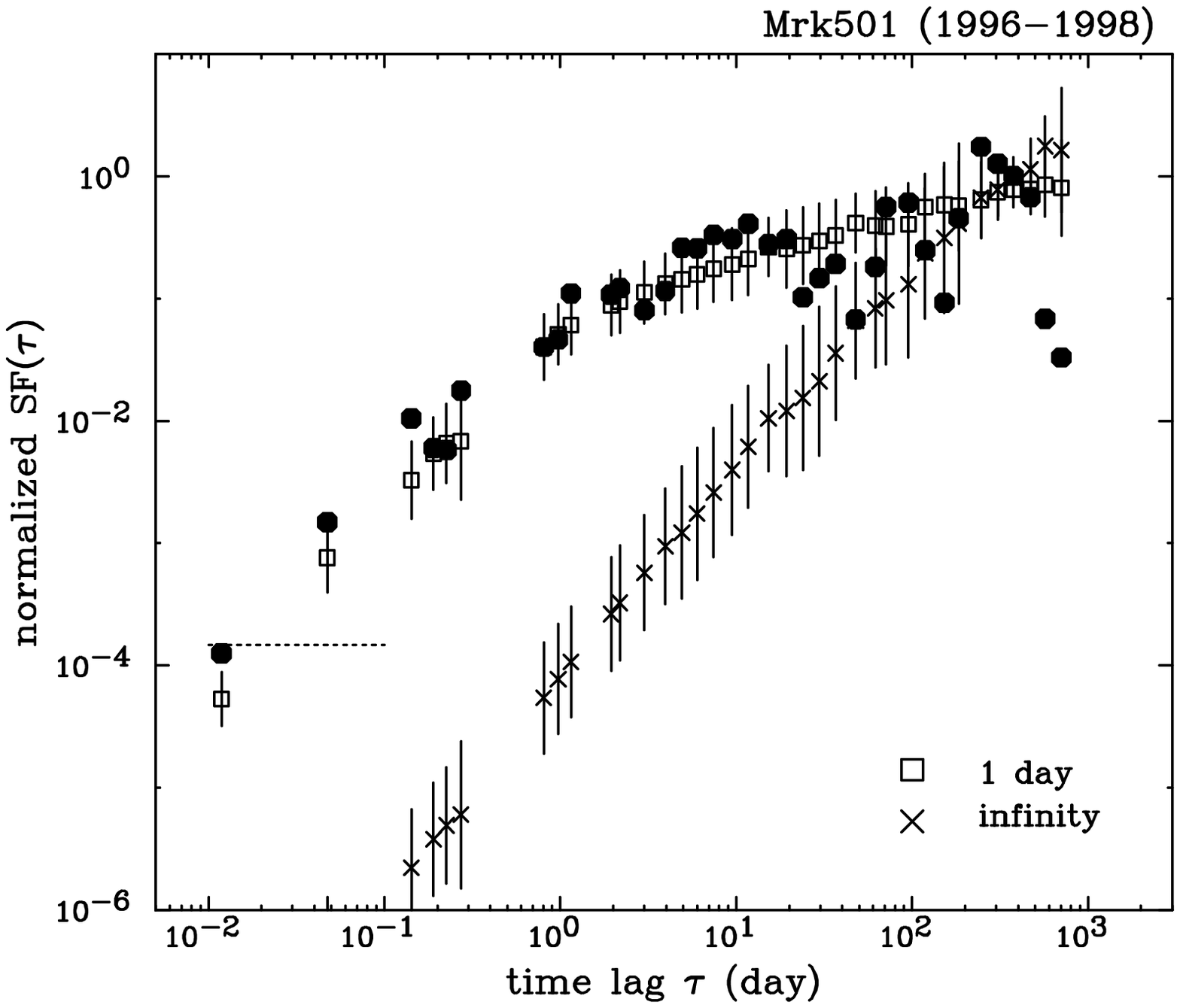}
\end{figure}
\clearpage

\begin{figure}
\epsscale{0.8}
\plotone{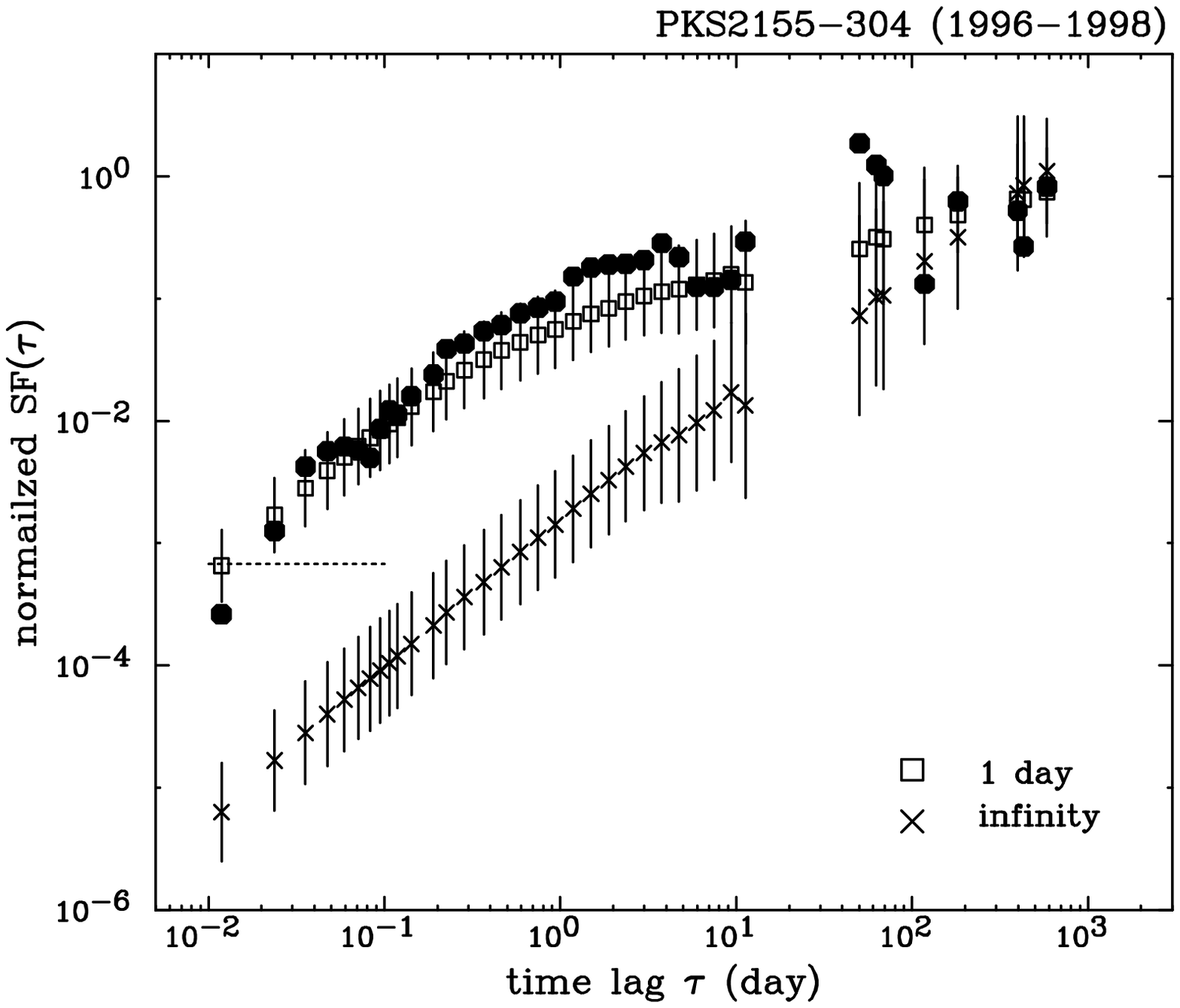}
\end{figure}
\clearpage

\begin{figure}
\epsscale{1.0}
\plotone{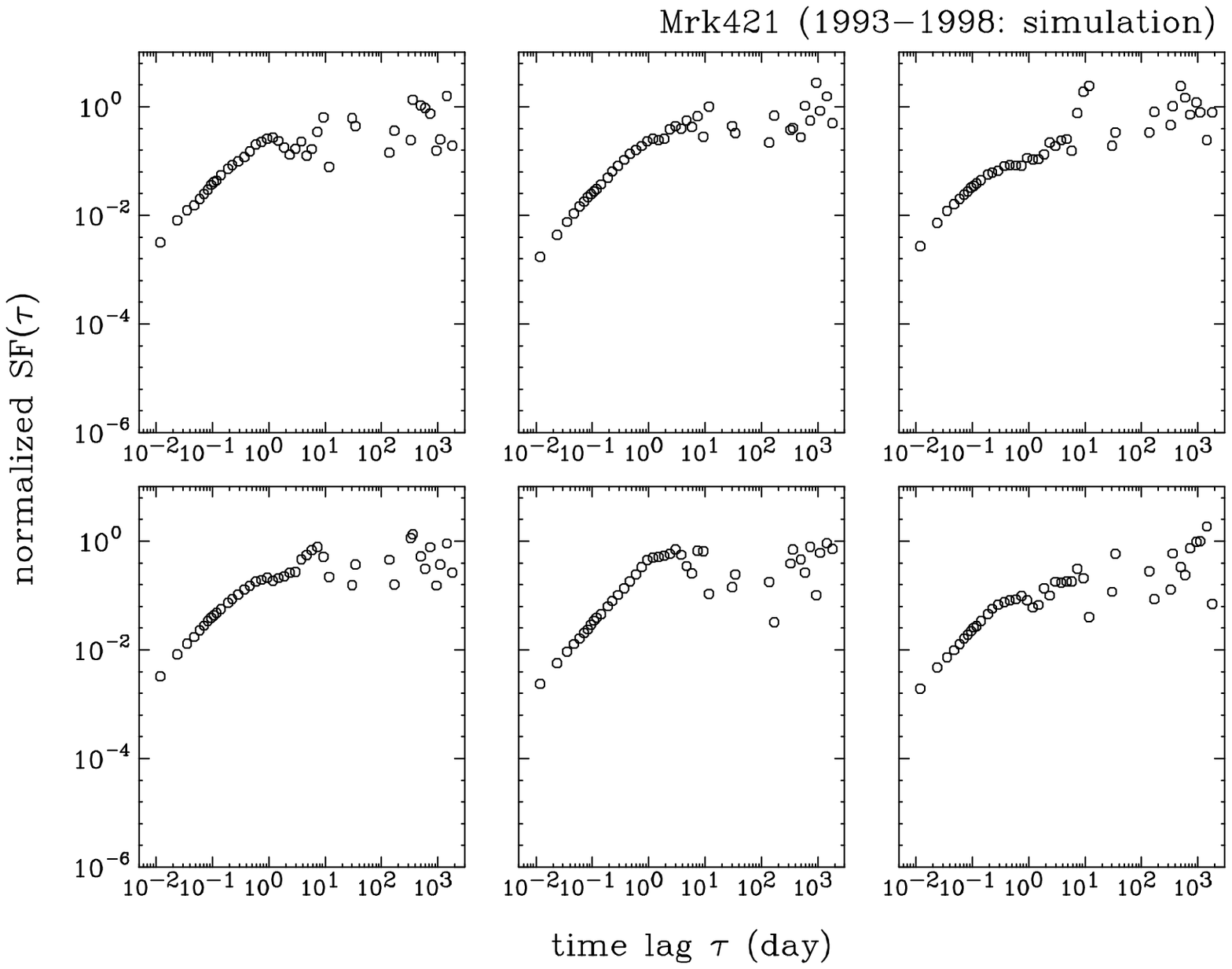}
\end{figure}
\clearpage

\begin{figure}
\epsscale{1.0}
\plotone{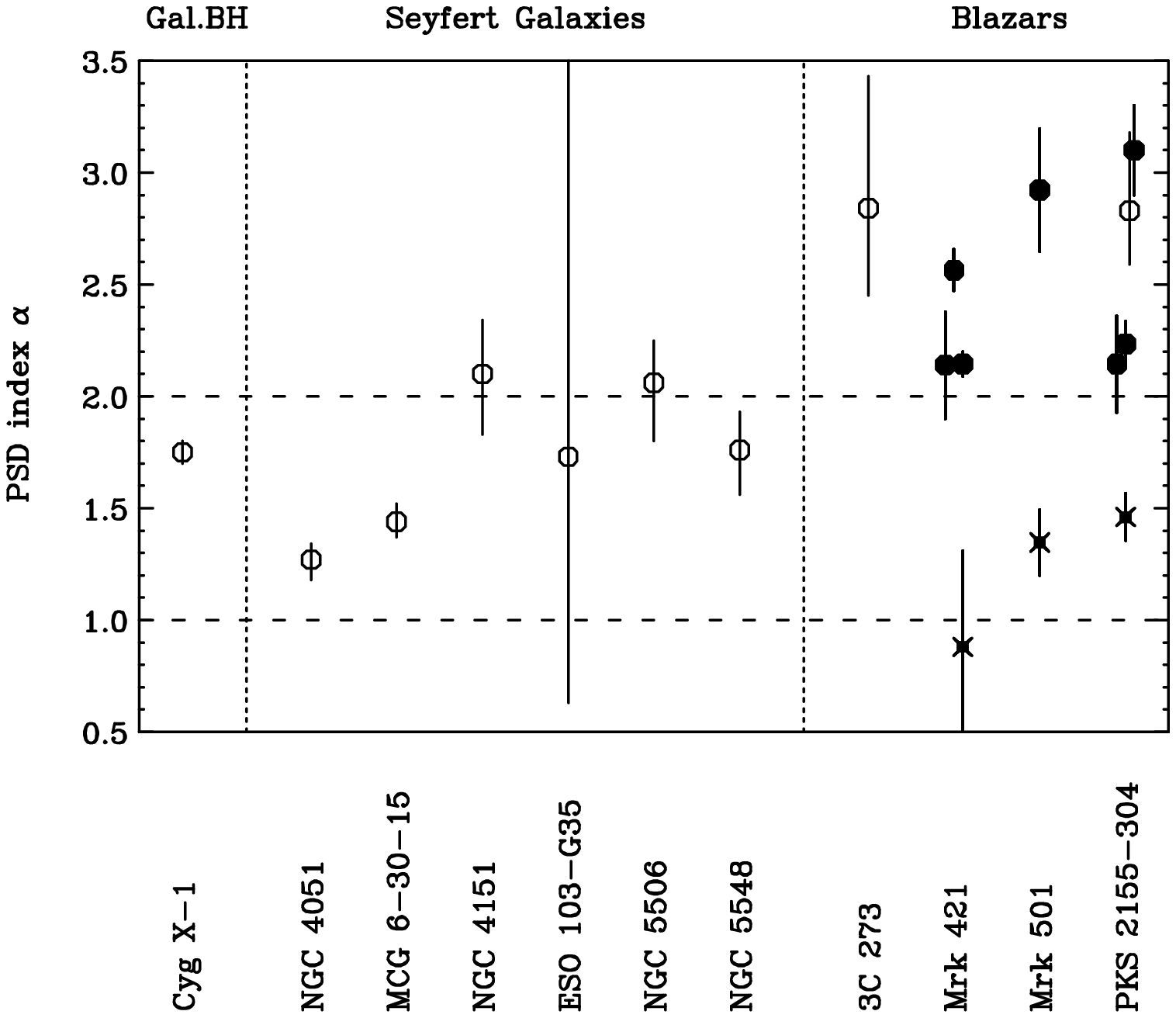}
\end{figure}

\clearpage

\begin{figure}
\epsscale{1.0}
\plotone{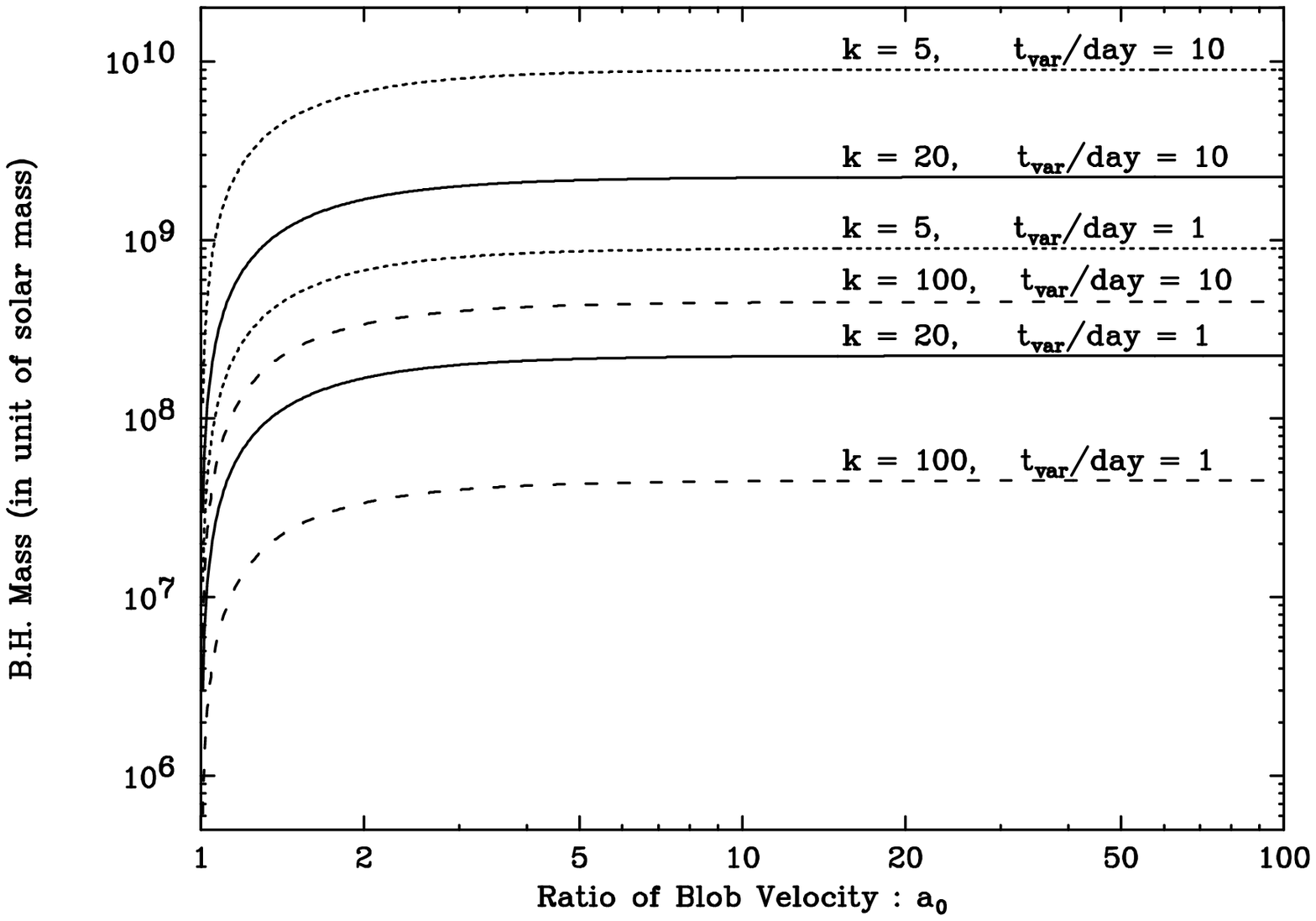}
\end{figure}
\clearpage

\end{document}